\documentclass[twocolumn]{aastex701}
\usepackage{apjfonts}
\usepackage{graphicx}
\usepackage{amsmath}
\usepackage{amssymb}
\usepackage{color}

\gdef\xx[#1]{\textcolor{red}{#1}}

\gdef\kms{km\,s$^{-1}$}
\gdef\msun{M$_{\odot}$}

\gdef\wake{RBH-1}

\gdef\lya{Ly\kern 0.09em$\alpha$}
\gdef\ha{H\kern 0.09em$\alpha$}

\newcommand{\GG}[1]{}

\lefthead{van Dokkum et al.}
\righthead{}
\begin{document}

\newcommand\XXX[1]{{\textcolor{red}{\textbf{x\ #1\ x}}}}

\title{JWST Confirmation of a Runaway Supermassive Black Hole via its Supersonic Bow Shock}


\author[0000-0002-8282-9888]{Pieter van Dokkum}
\affiliation{Astronomy Department, Yale University, 219 Prospect St,
New Haven, CT 06511, USA}
\affiliation{Dragonfly Focused Research Organization, 150 Washington Avenue, Suite 201, Santa Fe, NM 87501, USA}
\email{pieter.vandokkum@yale.edu}
\author{Connor Jennings}
\affiliation{Astronomy Department, Yale University, 219 Prospect St,
New Haven, CT 06511, USA}
\email{connor.jennings@yale.edu}
\author[0000-0002-7075-9931]{Imad Pasha}
\affiliation{Astronomy Department, Yale University, 219 Prospect St,
New Haven, CT 06511, USA}
\affiliation{Dragonfly Focused Research Organization, 150 Washington Avenue, Suite 201, Santa Fe, NM 87501, USA}
\email{imad.pasha@dragonfly1000.com}
\author[0000-0002-1590-8551]{Charlie Conroy}
\affiliation{Harvard-Smithsonian Center for Astrophysics, 60 Garden Street,
Cambridge, MA, USA}
\email{cconroy@cfa.harvard.edu}
\author[0009-0001-1399-2622]{Ish Kaul}
\affiliation{Department of Physics, University of California, Santa Barbara, CA 93106, USA}
\email{ikaul@ucsb.edu}
\author[0000-0002-4542-921X]{Roberto Abraham}
\affiliation{Department of Astronomy \& Astrophysics, University of Toronto,
50 St.\ George Street, Toronto, ON M5S 3H4, Canada}
\affiliation{Dragonfly Focused Research Organization, 150 Washington Avenue, Suite 201, Santa Fe, NM 87501, USA}
\email{roberto.abraham@utoronto.ca}
\author[0000-0002-1841-2252]{Shany Danieli}
\affiliation{School of Physics and Astronomy, Tel Aviv University, Tel Aviv 69978, Israel}
\email{sdanieli@tauex.tau.ac.il}
\author[0000-0003-2473-0369]{Aaron J.\ Romanowsky}
\affiliation{Department of Physics and Astronomy, San Jos\'e State University,
San Jose, CA 95192, USA}
\affiliation{Department of
Astronomy and Astrophysics, University of California Santa Cruz, 1156 High Street, Santa Cruz, CA 95064, USA}
\email{aaron.romanowsky@sjsu.edu}
\author[0000-0002-5445-5401]{Grant Tremblay}
\affiliation{Harvard-Smithsonian Center for Astrophysics, 60 Garden Street,
Cambridge, MA, USA}
\email{grant.tremblay@cfa.harvard.edu}


\begin{abstract}

We present JWST/NIRSpec IFU observations of a candidate runaway supermassive black hole
at the tip of a 62\,kpc-long linear feature
at $z=0.96$.
The JWST data show a sharp kinematic discontinuity at the tip, with
a radial velocity change of $\approx 600$\,\kms\ across $0\farcs 1$ ($1$\,kpc).
The velocity gradient, together with the projected post-shock flow velocity of
$\approx 300$\,\kms, 
is well described by a simple shock-compression model of a supersonic object,
with a velocity of 
$v_\bullet = 954^{+110}_{-126}$\,\kms\ and an inclination
$i=29^{+6}_{-3}$\,deg.
The previously puzzling kinematics along the linear feature, with the observed radial velocity
decreasing from $\approx 300$\,\kms\ near the tip to $\approx 100$\,\kms\ closer
to the former host galaxy, are naturally explained as gradual downstream mixing of shocked gas with the circumgalactic medium through turbulent entrainment. 
The bow shock interpretation is further supported by the morphology
of the gas at the tip of the wake and an analysis of the [O\,III]/H$\alpha$, [N\,II]/H$\alpha$,
[S\,II]/H$\alpha$, and [S\,III]/[S\,II] line ratios.
The line ratios are consistent with fast radiative shocks and rapid cooling, with best-fit
shock velocities that are in agreement with expectations from the black hole velocity and the shock geometry. Energy conservation over the lifetime of the wake suggests
a SMBH mass of $M_\bullet \gtrsim 10^7$\,\msun.
These results
confirm that the wake is powered by a supersonic runaway supermassive black hole, a long-predicted consequence of gravitational-wave recoil or multi-body ejection from galactic nuclei.

\end{abstract}


\section{Introduction}

The occasional escape of supermassive black holes (SMBHs) from their host galaxies is a long-standing prediction of theoretical studies. Velocity kicks can be imparted through two
distinct channels: a slingshot resulting from
a three-body interaction \citep{saslaw:74,volonteri:03,hoffman:07} or
gravitational wave recoil following a BH-BH merger \citep{bekenstein:73,campanelli:07}. 
Both channels occur naturally as a result of galaxy -- galaxy mergers, as the black
holes of the ancestor galaxies end up in the center of the descendant.
If the imparted velocity exceeds the escape velocity, the SMBH will leave
the host and continue to travel through intergalactic space \citep[see, e.g.,][]{saslaw:74,hoffman:07,lousto:12,ricarte:21}.

At present, no such ``runaway'' SMBHs have been confirmed, although there are
several promising candidates. Most of these are double or triple active galactic nuclei (AGNs),
or AGNs that are displaced from
the center of their host galaxy \citep[e.g.,][]{magain:05,komossa:08,civano:10,robinson:10,
chiaberge:17,liu:25}. The interpretation of these systems is often ambiguous; in particular, it
is difficult to distinguish ongoing mergers from ejections \citep[see, e.g.,][]{merritt:06}.
A special case is the $\infty$\,galaxy \citep{dokkum:25a}, also known as the  Cosmic Owl
\citep{li:25}. This object has two nuclei and three active SMBHs, with one of the
black holes embedded in a gas cloud
in between the two nuclei. This third SMBH might have escaped from
one of the nuclei in a three-body interaction, but in \citet{dokkum:25a} and
its JWST IFU follow-up \citet{dokkum:25b} it is argued
that it is more likely that it formed in-situ through a direct collapse.

Arguably the best candidate for a SMBH that has completely escaped from its former host is
a remarkable linear feature that was serendipitously discovered in an HST image
\citep[][hereafter Paper I]{dokkum:23}. The linear feature, here dubbed \wake,
extends 62\,kpc from a
$z=0.96$ galaxy. Keck/LRIS spectra show line emission along the entire extent of the
feature, culminating in a bright knot of [O\,III] emission at the tip. In Paper I we
interpreted the object as the wake of a runaway SMBH, with the tip the shock front
where the SMBH encounters the circumgalactic medium (CGM). Theoretical support for this interpretation
came from predictions of \citet{saslaw:72} and \citet{delafuente:08}, who studied the
interaction of runaway SMBHs with the surrounding gas. 

Following the publication of Paper I, 
\citet{ogiya:24} showed that
several of the observed features of \wake\ are indeed reproduced
in a hydrodynamical simulation
of the gravitational interaction between the CGM and a passing SMBH.
However, others have proposed alternative interpretations:
\citet{sanchez:23} suggest that the linear feature is
an edge-on, bulgeless galaxy, and \citet{chen:23} interpret the object as
a partially-shredded galaxy with a SMBH at one end.
The main limitation of the runaway SMBH interpretation was that it rested on circumstantial evidence: in Paper I we did not directly
demonstrate the presence of a high velocity object.
Here we present JWST/NIRSpec observations that provide
confirmation of a supersonic massive perturber at the tip of \wake.

\section{Data}

The key new data are JWST NIRSpec IFU observations. We also briefly describe new
HST imaging data and -- in \S\,\ref{downstream.sec} and \S\,\ref{dispersion.sec} --
re-analyze the Keck/LRIS data of Paper I.

\subsection{Deep HST WFC3/UVIS Imaging}
\label{uvis.sec}

The field of \wake\ was observed with HST WFC3/UVIS in Cycle 30, in program
HST-GO-17301 (PI: van Dokkum).
Two long pass filters were used, F200LP and F350LP, with six orbits in each of the filters.
Three exposures were obtained in each orbit, with small offsets to facilitate
bad pixel removal and improve subpixel sampling. The total exposure time is 14,976\,s in
F200LP and 14,922\,s in F350LP. The main goal of the observations was to study
the rest-frame far-UV emission of \wake: the difference between the F200LP and
F350LP filters, ${\rm LP}_{\rm diff}$, creates a broad UV filter with central rest-frame wavelength
$\lambda_{\rm rest} \sim 1400$\,\AA.

The data were reduced in the following way. In the UVIS longpass filters the background
shows large scale gradients, caused by the difficulty of flat fielding these very broad
filters \citep[see][]{mack:16}. We used a low order polynomial fit to remove the
background for each individual chip in each exposure.
The F200LP and F350LP exposures were then drizzled to an $0\farcs 04$\,pix$^{-1}$
grid at an orientation of 122$^{\circ}$. This orientation places
the linear feature along the $x$-axis with the galaxy at left and the tip at right.
The ${\rm LP}_{\rm diff}$ image was constructed by the subtraction
${\rm LP}_{\rm diff} = {\rm F200LP} - \kappa {\rm F350LP}$, where $\kappa$
encodes the differences in response between the filters at $\lambda > 3500$\,\AA.
We determined $\kappa \approx 1.22$ using early-type galaxies in the field.
We also created the sum of the filters, ${\rm LP}_{\rm sum}=
{\rm F200LP} + {\rm F350LP}$, which is effectively
a 12-orbit deep ``white light'' optical image. 

\begin{figure}[ht]
  \begin{center}
  \includegraphics[width=0.95\linewidth]{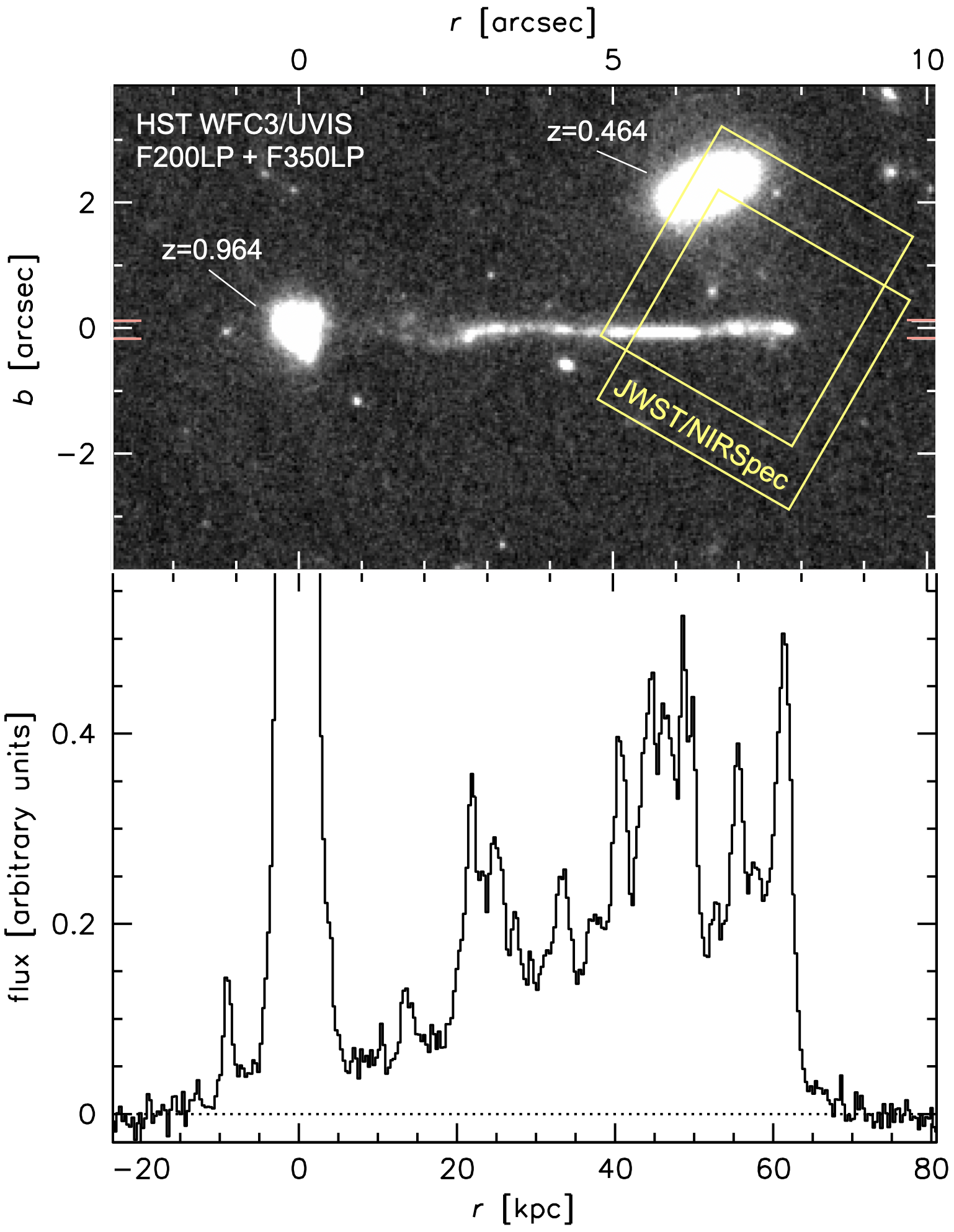}
  \end{center}
\vspace{-0.2cm}
    \caption{
{\em Top:} HST/WFC3 UVIS imaging of the candidate $z=0.96$ runaway black hole wake \wake.
The displayed image spans $13\farcs 2 \times 7\farcs 7$.
It is the sum of two long pass filters, F200LP and F350LP, with a combined integration
time of 29,898\,s. The two JWST NIRSpec pointings are indicated in yellow.
{\em Bottom:} summed flux profile along the feature. There is
continuous emission all the way from the galaxy to the tip at $r\approx 62$\,kpc,
followed by a sudden drop of a factor of $>40$.
The profile does not resemble the exponential fall-off
expected for an edge-on spiral galaxy \citep[an
interpretation that was proposed by][]{sanchez:23}.
}
\label{uvis.fig}
\end{figure}

The ${\rm LP}_{\rm sum}$ image of the linear feature and the associated
galaxy is shown in Fig.\ \ref{uvis.fig}. The intensity along the feature is shown in
the panel below the image, derived from summing the central 8 rows ($0\farcs 32$).
The deep data show that there is continuous emission all the way from the
galaxy to the tip of the feature, including in the apparent gap between the
streak in the original ACS-imaging and the galaxy \citep[something we also noted in VLT $B$ band data; see][]{dokkum:23rnaas}. Furthermore, the data confirm that the feature does not extend
beyond the tip at $r\approx 62$\,kpc. The drop is a factor of $>40$ at 95\,\% confidence,
which is difficult to reconcile with the expected exponential decline of an
edge-on disk or the surface brightness profile of a tidal feature \citep{sanchez:23,chen:23}.

\begin{figure*}[ht]
  \begin{center}
  \includegraphics[width=0.8\linewidth]{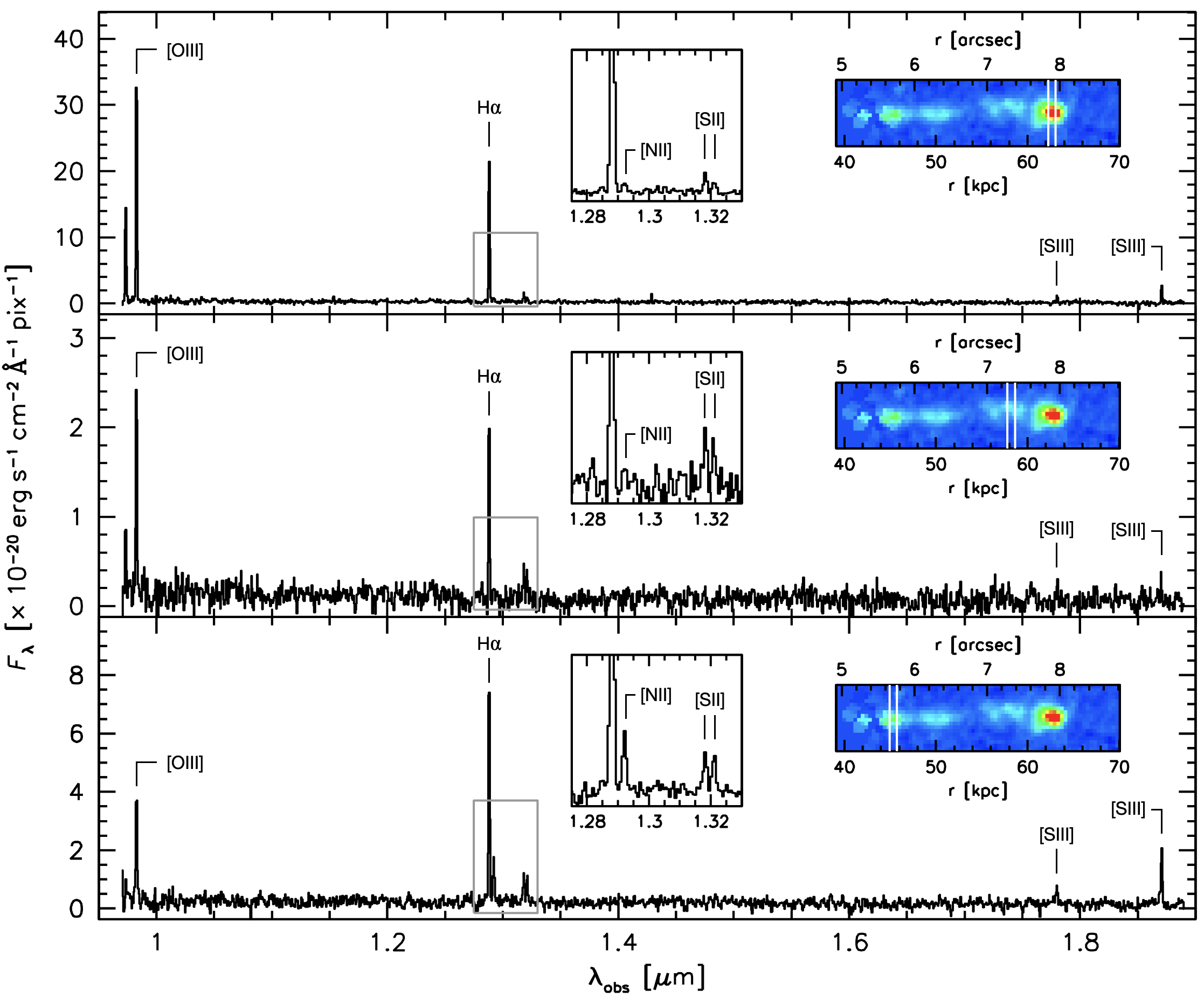}
  \end{center}
\vspace{-0.2cm}
    \caption{
Overview of the NIRSpec IFU data. Insets show a map of the redshifted [O\,III] emission,
with three characteristic locations indicated by vertical lines. The main
panels show the spectra at those locations. There is a distinct, very bright
knot at the tip of \wake, with a high [O\,III]/H$\alpha$ ratio. This
ratio decreases gradually behind the tip, while the [N\,II]/H$\alpha$ and
[S\,II]/H$\alpha$ ratios increase. In all panels [O\,III] and H$\alpha$ are the
strongest lines; these are used for the dynamical analysis in \S\,\ref{kinematics.sec}.
}
\label{overview.fig}
\end{figure*}

The data do not show any emission on the opposite side of the galaxy, in either
${\rm LP}_{\rm sum}$ or ${\rm LP}_{\rm diff}$, in conflict with a low significance CFHT $u$-band
detection of a counter-wake
in Paper I. One of the goals of the UVIS observations was to assess whether
this apparent counter-wake actually exists, and we conclude that it does not \citep[see also][]{montes:24}.

\subsection{JWST NIRSpec Observations}

\subsubsection{Observing Program and Data Reduction}

We obtained spatially-resolved spectroscopy of the tip of \wake\ with the JWST NIRSpec Integral
Field Unit (IFU) on July 24 2024, in Cycle 2 program JWST-GO-3149 (PI: van Dokkum).
The observations are divided over two target positions, symmetrically placed on each side of the linear feature $0\farcs 5$ downstream from the tip (see Fig.\ \ref{uvis.fig}). 
The $0\farcs 1$ spaxels of the IFU undersample the point spread function (PSF);
at each of the two target positions a 4-point dither pattern samples the PSF
on a $0\farcs 05$ grid. The G140M/F100LP grating and filter combination gives
wavelength coverage from 0.97\,$\mu$m\,--\,1.89\,$\mu$m, corresponding to
0.49\,$\mu$m\,--\,0.96\,$\mu$m in the restframe. The spectra do not include H$\beta$
but cover [O\,III], H$\alpha$, [N\,II], [S\,II], and [S\,III], along with fainter lines.
The spectra are sampled on a grid with a $6.4$\,\AA\ spacing, corresponding to 193\,\kms\ at
[O\,III] and 147\,\kms\ at H$\alpha$. The total integration time was 3549\,s for each pointing,
for a total of 7089\,s.

\begin{figure*}[ht]
  \begin{center}
  \includegraphics[width=1.0\linewidth]{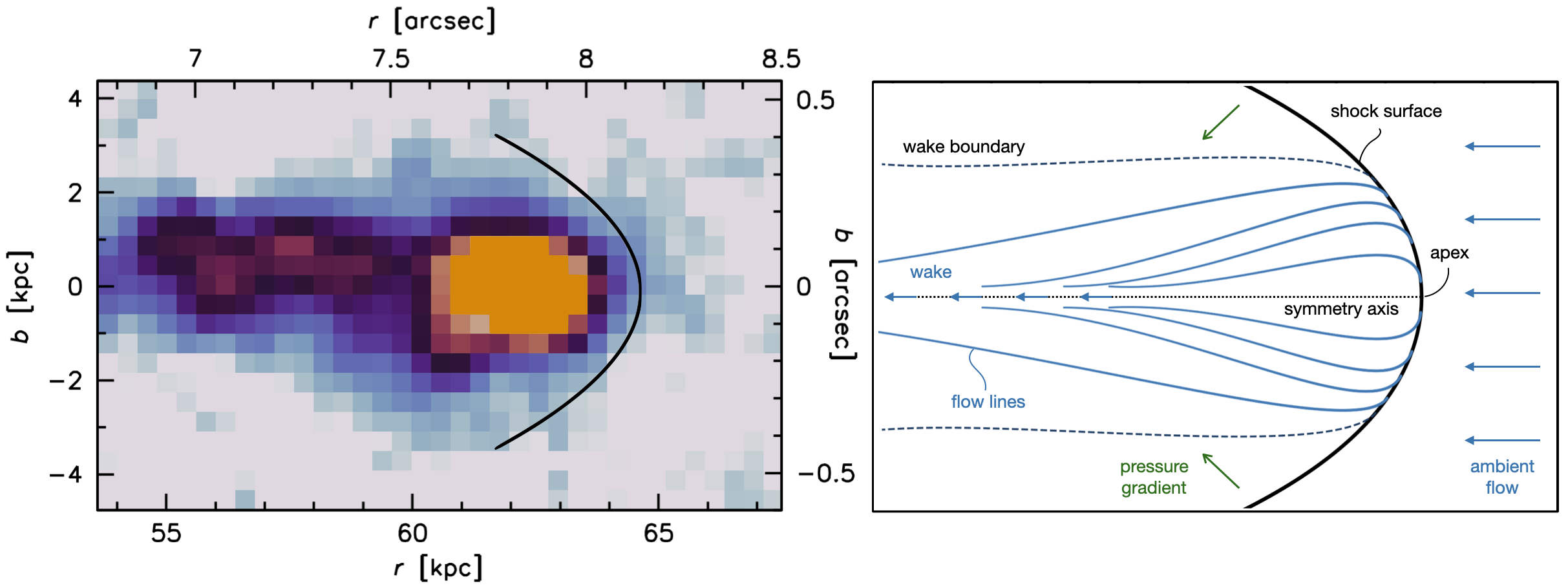}
  \end{center}
\vspace{-0.2cm}
    \caption{
{\it Left:} Summed [O\,III]\,+\,H$\alpha$ line emission at the tip of the
linear feature.
The tip is clearly resolved, and resembles a classic bow shock with a
thin turbulent
wake behind it.
The thick black line is a  restricted parabolic fit to a
faint isophote at $63\leq r \leq 65$\,kpc.
{\em Right:} Illustration of a bow shock, at approximately the same
scale. Upon encountering the shock gas flows tangentially along the shock
surface. The low pressure behind the shock causes flow lines to bend
toward the symmetry axis, creating a narrow wake.
}
\label{apex.fig}
\end{figure*}

The data were reduced with the TEMPLATES pipeline \citep{TEMPLATES}, built on the JWST Science Calibration Pipeline \citep{JWSTpipe} (v1.18.0). The method follows that of \citet{dokkum:25b}; here we summarize briefly. Uncalibrated exposures from MAST (JWST SDP v2025\_2) were processed through the stage 1 detector-level pipeline, followed by $1/f$ noise correction with NSClean \citep{rauscher:24}. Stage 2 spectroscopic calibration was run, and cosmic rays were identified as outlier spaxels.
The NRS1 data from all exposures and all dithers were then combined, with no background
subtraction.  Taking
advantage of the subpixel sampling of the dither pattern, the
data were combined onto a $0\farcs 05$ grid, corresponding to the angular scale that is sampled by the dither pattern. We note that the G140M/F100LP data cubes show no obvious ``wiggles'', low frequency sinusoidal artifacts of undersampling
that are often seen in NIRSpec IFU data \citep[see, e.g.,][]{perna:23,dumont:25}.

These steps were performed for each of the two target positions. To remove background structures and
detector artifacts,
we subtracted the two pointings from one another. A copy of the difference cube was shifted by the
distance between the two pointings, multiplied by $-1$, and added to the original difference cube.
This procedure results in a positive summed version of the linear feature
in the center of the field, flanked by negative
versions from each of the two pointings.

\subsubsection{Qualitative Description of the Data}

An overview of the reduced data is shown in Fig.\ \ref{overview.fig}. 
The insets show
a map of the [O\,III] emission, created by summing the three wavelength bins with strong
[O\,III] emission. There is significant structure in the feature. The tip is distinct
from the rest of \wake\
and very bright in [O\,III]. There is no detected emission upstream from the tip.
While the tip is compact, it is spatially resolved (as 
anticipated in paper I). Behind the tip, at 55\,kpc\,$\lesssim r\lesssim 60$\,kpc, is a
turbulent region with several knots. Further downstream the
wake narrows and there are more concentrations.

Three spectra
are shown, for the tip, the turbulent region just behind it, and an
area further downstream. The strongest
redshifted emission
lines are [O\,III]\,$\lambda 5007$, H$\alpha,$ [N\,II]\,$\lambda 6584$, the
[S\,II]\,$\lambda\lambda 6716,6731$ doublet, and the [S\,III]\,$\lambda\lambda 9069,9532$
doublet.
There are striking differences between the spectra. At the tip, the strongest line
is [O\,III], [N\,II] is barely detected, and [S\,II] is faint compared to [S\,III].
Behind the tip the [O\,III]/H$\alpha$ ratio decreases and [S\,II]/H$\alpha$ ratio
increases. The [N\,II] line remains weak, until we are quite far downstream.
We will return to this in \S\,\ref{line_ratios.sec}.


\section{Morphological Evidence for a Bow Shock with a Turbulent Wake}
\label{morph.sec}

We begin by analyzing the ionized gas
morphology of the tip region in the JWST data cube. In paper I we predicted that
the shock should be spatially-resolved at $\sim 0\farcs 1$ resolution: the 
[O\,III] surface brightness of fast radiative shock is limited by the
incoming energy flux, and the
models of \citet{dopita:96} imply a total area of the shock front of
$\sim 0.2n^{-1}$\,kpc$^2$ for the observed [O\,III] luminosity (with $n\sim
10^{-3}$ the density of the CGM).


The tip of the wake is shown in Fig.\ \ref{apex.fig}.
To create this image we summed the [O\,III]\,$\lambda 5007$ and H$\alpha$ maps in order
to maximize the S/N ratio of faint emission. 
As expected the tip region is not a point source
but clearly resolved, both in the direction
away from the galaxy
($r$) and in the perpendicular ($b$) direction. The faint contours flare
out from the tip to reach a total width of $\Delta b \approx 6$\,kpc at
$r\approx 61$\,kpc and then taper downstream.  The emission breaks
up into several distinct knots at 55\,kpc\,$\lesssim r \lesssim$\,60\,kpc.

This is the classic morphology of a bow shock with a turbulent wake,
as illustrated in the right panel of Fig.\ \ref{apex.fig}. The ambient
flow, here the CGM, is decelerated in the direction normal to the
shock and deflected along the shock surface.  The low pressure behind
the shock front then bends the flow lines toward the symmetry axis, leading
to tapering and a narrow wake \citep[see, e.g.,][]{Wilkin1996,bucciantini:02,
tarango:18,meyer:17}.
Well-known examples
with similar morphology are the bow shocks associated with Mira \citep{martin:07}
and the Herbig-Haro objects HH\,1 \citep{bally:02} and HH\,34 \citep{reipurth:02}.

It is not very meaningful to quantify the width of the shock with traditional
measures such as the FWHM: the emission
lines do not peak in the shock front itself but in the compression
and cooling zone behind it,
as this is where the temperature has decreased to $10^{4-5}$\,K and
the density is highest. Flux-weighted quantities therefore measure
the extent of the cooling zone, not the shock front.
Instead, we characterize the geometry by fitting a parabola to an outer contour.
The local radius of curvature at the apex, $R_{\rm c}$, is relatively robust against
thresholds or the flux distribution inside the shock. Once $R_{\rm c}$ is known, the shock radius downstream follows from the parabolic fit, using
$R_{\rm phot}(\Delta r_{\rm apex}) = \sqrt{2\,R_{\rm c}\,\Delta r_{\rm apex}}$ with $\Delta r_{\rm apex}$ the downstream distance from the location of the apex.

The thick line in Fig.\ \ref{apex.fig}
shows the best-fitting parabola, with the fit restricted to
a faint contour ($1\sigma$ above the background)
and a radial range 63\,kpc\,$\leq r\leq$\,65\,kpc.
The apex location of the parabola
$r_{\rm apex} \approx 64.6$\,kpc (that is, 64.6\,kpc from the galaxy)
and the radius of curvature  $R_{\rm c} \approx 1.8$\,kpc.
The associated stand-off scale is
$R_0 \approx \frac{2}{3} R_{\rm c} \approx 1.2$\,kpc \citep{Wilkin1996}.
The expected location of the black hole is therefore $\approx 1.2$\,kpc
downstream of the apex, at $r \approx 63$\,kpc. We return to this in \S\,\ref{blackhole.sec}. The shock extends approximately 4\,kpc behind the apex and its total
area is $\sim 80$\,kpc$^2$, in reasonably good agreement with our
[O\,III] luminosity-based estimate in Paper I.

\section{Bow Shock Kinematics}
\label{kinematics.sec}


\subsection{Qualitative Expectations}

Bow shocks produce distinct signatures in the line-of-sight (LOS)
kinematics that can be used to
infer the orientation and velocity of the shock. As discussed above
and shown in Fig.\ \ref{apex.fig}, gas initially
flows along the limbs of the bow and then bends towards the symmetry axis
to form the wake. If the shock is viewed exactly side-on ($i=0^{\circ}$ with
respect to the plane of the sky), there will be no
LOS velocity in the wake region: all the motion in the wake is in the plane of the sky.
The velocities in the limbs of the bow are strongly redshifted and blueshifted,
as gas on the near limb is coming toward the observer and gas on the far limb
is moving away from the observer. The net effect is a broad LOS velocity distribution
at the tip and a narrow distribution further downstream, with a mean velocity of $\sim 0$
everywhere. 
If the shock is viewed exactly from the front ($i=90^{\circ}$)
the LOS velocity will be dominated
by the wake, as the gas on the limbs flows mainly in the plane of the sky. The
wake velocity will be blueshifted with respect to the ambient medium if
the object is coming towards the observer and redshifted if it is moving
away from the observer.

\begin{figure}[ht]
  \begin{center}
  \includegraphics[width=0.95\linewidth]{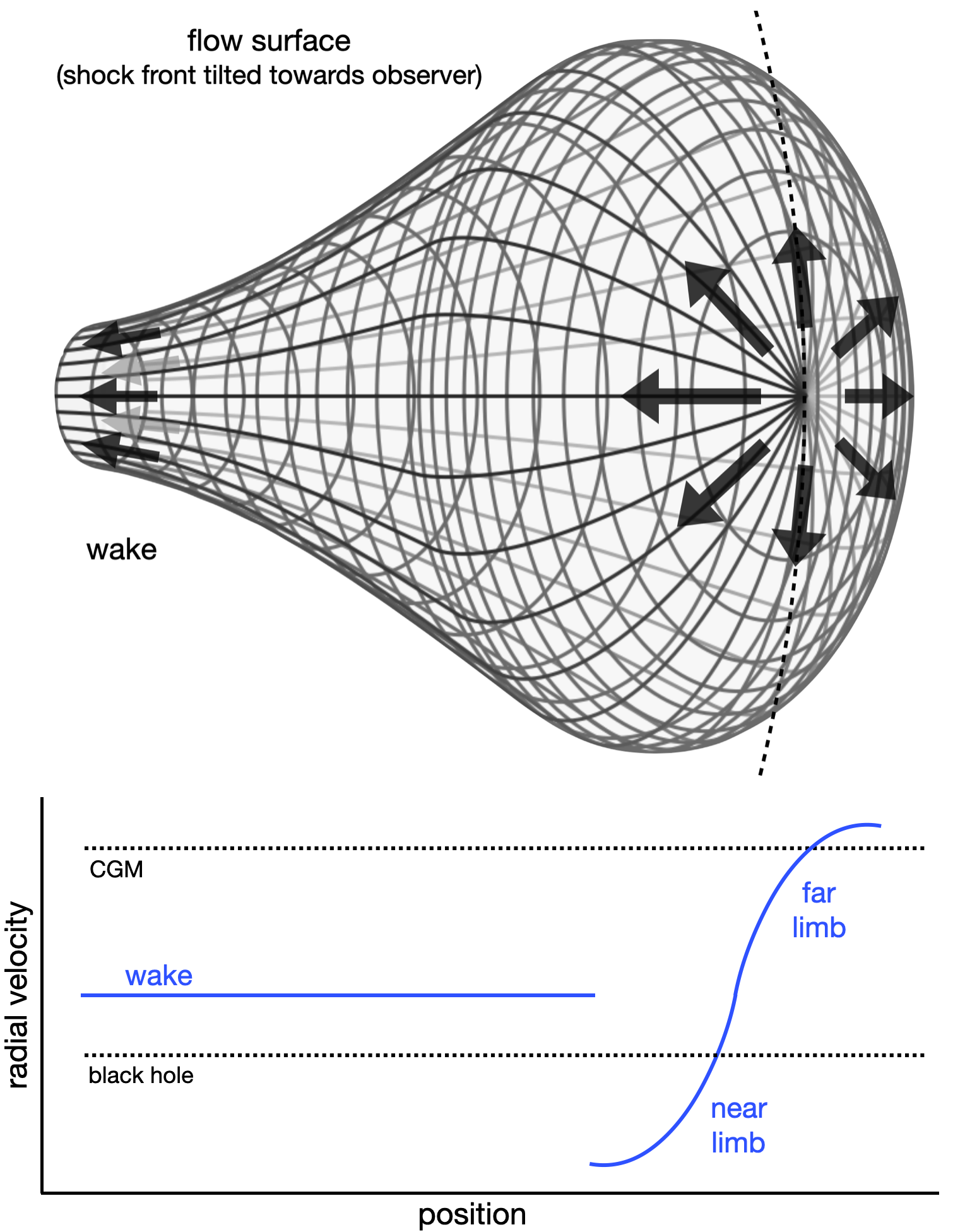}
  \end{center}
\vspace{-0.2cm}
    \caption{
Illustration of a flow surface for a shock that is projected at an intermediate
angle ($50^{\circ}$) toward the observer. Gas on the near limb is blueshifted
and gas on the far limb is redshifted with respect to the wake velocity.
The blueshifted gas is projected further downstream than the redshifted
gas, leading to a velocity gradient.
}
\label{mushroom.fig}
\end{figure}

More commonly the shock will be tilted at an intermediate angle. This situation
is illustrated in Fig.\ \ref{mushroom.fig}, where we show a 3D rendering
of a flow surface that is tilted toward the
observer at an angle of 50$^{\circ}$.
With respect to the LOS velocity in the wake,
gas on the near limb is blueshifted and gas on the far limb is redshifted.
Furthermore, the blueshifted limb is projected at a different location than
the redshifted limb: going downstream from the apex, the gas is first
redshifted and then blueshifted. As a result, bow shocks often produce
a tilted locus in position-velocity
diagrams for intermediate viewing
angles \citep[see, e.g.,][]{hartigan:87,indebetouw:95,lopez:02}.

\begin{figure*}[ht]
  \begin{center}
  \includegraphics[width=1.0\linewidth]{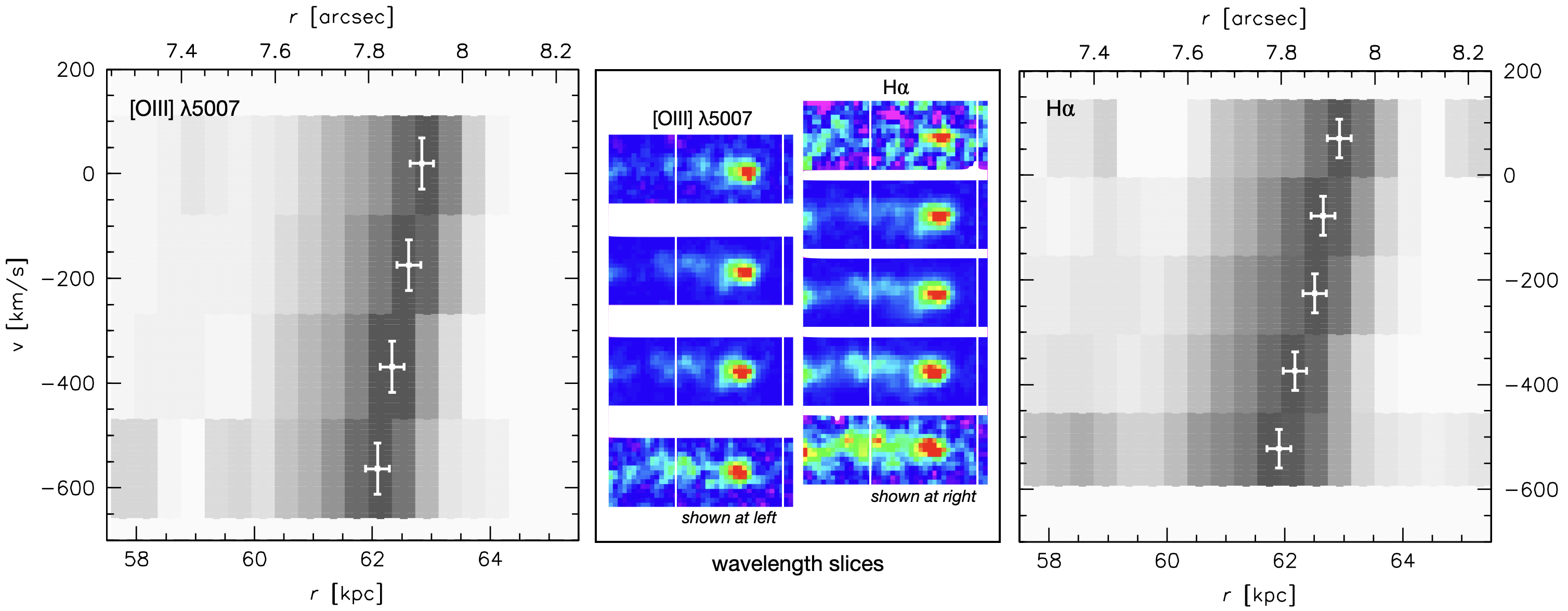}
  \end{center}
\vspace{-0.2cm}
    \caption{
{\em Middle:} Wavelength slices through the data cube, near the wavelength of the redshifted
[O\,III]\,$\lambda 5007$ line (left) and H$\alpha$ (right). Fluxes have been scaled to
the peak in each slice. There is a striking pattern, with the emission near the
tip systematically shifting further upstream for higher velocities.
{\em Left and right:} Position-velocity diagrams for the regions in between the
white lines, obtained by averaging the flux in the direction perpendicular to the wake. 
These diagrams show an unambiguous gradient of $\sim 600$\,\kms\ over $\sim 1$\,kpc.
The gradient is quantified by the white data points.
}
\label{pv.fig}
\end{figure*}

\subsection{A Steep Velocity Gradient at the Tip of the Wake}
\label{gradient.sec}

Remarkably, inspection of the data cubes shows clear evidence for such a gradient
at the tip of \wake:
the centroid of emission moves systematically upstream with increasing
wavelength. This is demonstrated in the central panel of Fig.\ \ref{pv.fig}, 
where we show wavelength channels around the redshifted [O\,III]\,$\lambda 5007$
and H$\alpha$ lines. By collapsing the flux in these maps
vertically, that is, perpendicular to the wake direction, we obtain
position-velocity diagrams, shown in the left and right panels of Fig.\ \ref{pv.fig}.
Following the convention of Paper I we calculate
velocities with
respect to the redshift of the galaxy. 

We find a well-defined and extremely steep velocity gradient in both [O\,III] and H$\alpha$,
going from about $-600$\,\kms\ to $+50$\,\kms\ across $\sim 1$\,kpc ($0\farcs 1$).
The trends are quantified with a quadratic interpolation to find the location of the centroid in each velocity channel (white data points).
We note that the {\em peak} of emission is always in the central channels, that is,
near $\sim -300$\,\kms. This is consistent with the Keck spectroscopy of Paper I,
which shows a plateau at that approximate velocity beyond $r\gtrsim 50$\,kpc.

\begin{figure*}[ht]
  \begin{center}
  \includegraphics[width=0.93\linewidth]{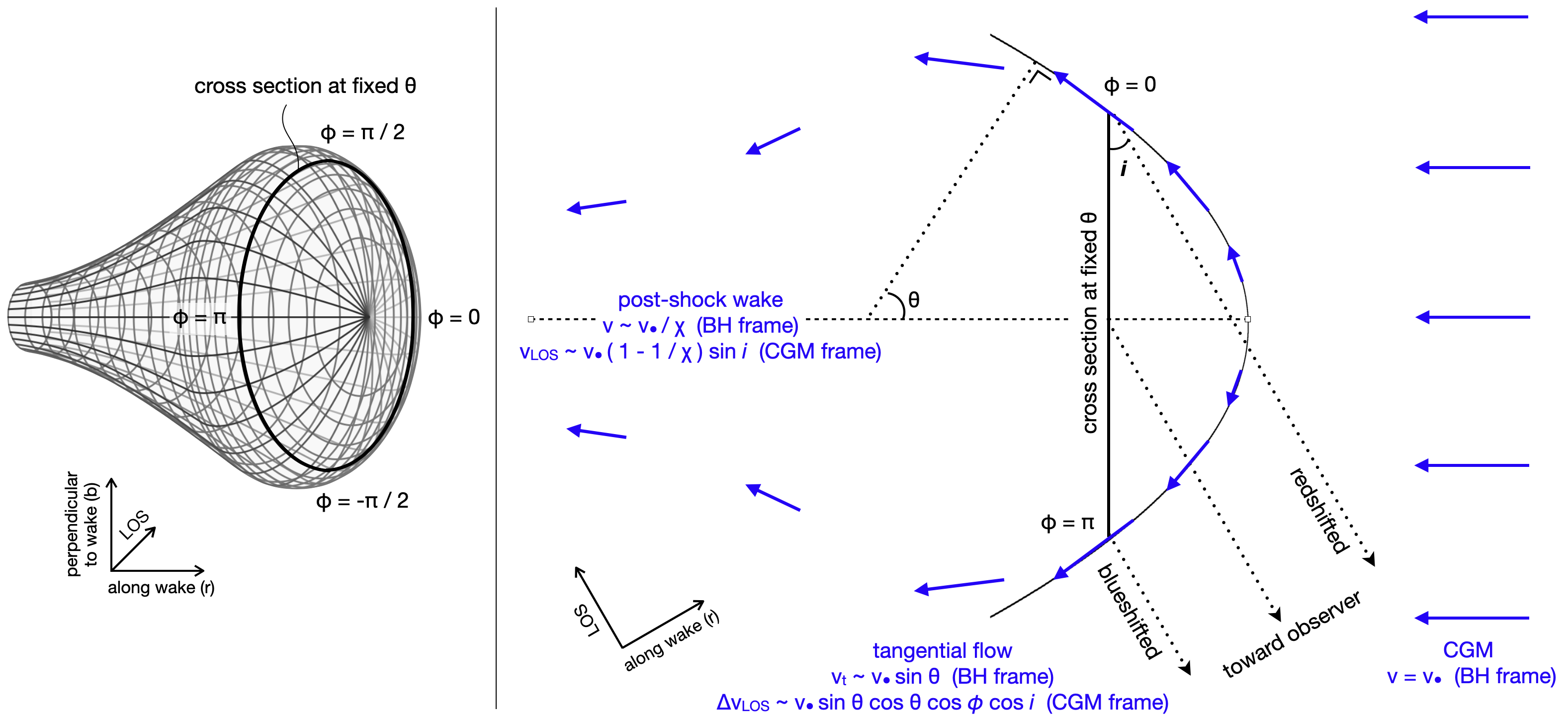}
  \end{center}
\vspace{-0.2cm}
    \caption{
Schematic of our bow shock model, broadly
following \citet{Wilkin1996}. {\em Left:} 3D representation of a flow
surface, as in Fig.\ \ref{mushroom.fig}. The flow on the shock surface
is modeled as a cross-section at a fixed opening angle
$\theta$. {\em Right:} 2D projection.
Gas flows, in the rest-frame of the black hole,
are indicated with blue arrows. The LOS velocity, in the rest-frame
of the CGM, is modeled as the sum of the
post-shock wake (which produces the plateau velocity downstream) and the tangential flow. The tangential flow produces a blueshifted
component from the near limb and a redshifted component from the far limb.
}
\label{umbrella.fig}
\end{figure*}

We thus find that the observed kinematics at the tip of \wake\ are qualitatively consistent
with expectations for a strong supersonic bow shock. Furthermore,
in Appendix \ref{grav.sec} we show that
the observed gradient can {\em not} be produced by the gravitational influence of a black hole,
as the implied black hole mass would be orders of magnitude higher than is reasonable.
In the following, we develop
a quantitative bow shock
model to determine key parameters such as the velocity of the black hole
and the inclination of the wake.

\subsection{Quantitative Bow Shock Model}

\subsubsection{Velocity Components}
\label{vkin.sec}

The model is a simplified version of the full treatment offered by
\citet{Wilkin1996} and others, geared toward fitting the velocity gradient.
The geometry and relevant gas flow directions are illustrated in Fig.\ \ref{umbrella.fig}.
While the gas flow is continuous, we separate it into two distinct components:
the tangential flow on the shock surface and the flow in the post-shock wake.
The tangential flow dominates near the apex, and is to good approximation
\begin{equation}
v_t(\theta) \approx v_\bullet\sin\theta,
\end{equation}
with $\theta$ the local shock obliquity. This is the angle
between the shell normal
and the black hole direction of motion, as indicated in Fig.\ \ref{umbrella.fig}:
$\theta=0^\circ$ at the apex and increases downstream as the shock surface flares. 

The flow in the wake is dominated by the velocity component that is normal
to the shock surface. The shock reduces this normal component
by  the compression factor $\chi$, such that 
\begin{equation}
v_{\rm wake} \approx v_n \approx \frac{v_\bullet}{\chi}.
\end{equation}
The compression factor satisfies $\chi\le 4$ for a strong adiabatic shock with $\gamma=5/3$ \citep[e.g.,][]{Draine2011}, and in typical ISM conditions the immediate compression is typically $\chi=3$--$3.5$ \citep[e.g.,][]{Anderson2003,DraineMcKee1993,Draine2011}. We  adopt $\chi=3$ in the main analysis.

These expressions are in the rest-frame of the black hole, that is, the flow is
seen as coming towards a static shock. However, 
the observed kinematics are not in the black hole rest-frame but in the rest-frame
of the CGM (or, more precisely, that of the former host galaxy), and also projected along
the line of sight. Adding the black hole velocity $+v_\bullet$ and
tilting the symmetry axis by the
inclination $i$ (the angle out of the sky plane, toward the observer) gives
the following expression for the normal component:
\begin{equation}
v_{{\rm LOS},\,{\rm wake}}
\approx v_\bullet\!\left(1-\frac{1}{\chi}\right)\!\sin i .
\label{eq:wake_los}
\end{equation}
The entire structure acquires this baseline in the BH\,$\rightarrow$\,CGM translation.

Next, we project the tangential velocity along the line of sight.
For simplicity we determine the line of sight velocity
for a circular cross section at a fixed obliquity $\theta$.
The azimuth around the axis of the cross section
is denoted by $\phi$, with the convention that $\phi=0$ at the far limb (back, away
from the observer),
$\phi=\pi$ at the near limb (front, toward the observer),
and $\phi=\pm \pi/2$ at the sides.
Projecting the tangential direction to the LOS gives the near–far (antisymmetric)
contribution
\begin{equation}
\Delta v_{\rm LOS}(\phi;\theta)
\approx v_\bullet \sin\theta\,\cos\theta\,\cos i\,\cos\phi .
\label{eq:dvl}
\end{equation}
Hence the observed LOS velocity in the region near the apex is
\begin{equation}
v_{\rm LOS}(\phi;\theta)\approx v_{{\rm LOS},\,{\rm wake}}\;\pm\;\Delta v_{\rm LOS},
\end{equation}
with the sign set by the local $\cos\phi$ (blueshifted on the near limb, redshifted
on the far limb).
Downstream from the shock surface only the wake baseline,
Eq.~\ref{eq:wake_los}, is present.

While simplified, this model
(tangential flow on a near-apex paraboloid shell) follows the
standard framework for the kinematics of
bow shocks \citep{Wilkin1996,Hartigan1987}.

\subsubsection{Turning the 2D Model into a 1D Model}

The velocity gradient is measured from spectra collapsed along the projected coordinate
\(b\) (perpendicular to the wake). We therefore define
$x\equiv r-r_0$, with $r_0$ the center position of the projected
cross section along the wake, and average the emission in the
model over the chord through the cross section.
Inside the region where the flow is on the
shock surface ($|x|<a$, with $a$ the projected half-width of the
cross section) the LOS–averaged azimuthal factor is
\begin{equation}
\langle\cos\phi\rangle(\xi)=\dfrac{\xi}{\sqrt{1-\xi^{2}}}\,
\mathrm{asinh}\!\big(\dfrac{\sqrt{1-\xi^{2}}}{|\xi|}\big),
\end{equation} 
with $\xi\equiv x/a$. This expression encodes the sign flip of the velocities
as the near and far sides of the cross section are projected along the line of sight:
for $-a<x<0$ the azimuthal factor is negative and for $0<x<a$ the factor is
positive.

In the 2D\,$\rightarrow$\,1D projection we also need to account for the
spatial distribution of emission within the cross section.
Bow shocks are often limb–brightened: the rims
contribute more emission than the interior \citep[e.g.,][]{Hartigan1987}. 
To implement rim–brightening we assume a simple
local emissivity law within a circular cross section of projected radius
$a$:
\begin{equation}
\label{limb.eq}
\epsilon\,(d_{\rm cross})\ \propto\ \left(\frac{d_{\rm cross}}{a}\right)^{p}, \qquad 0\le d_{\rm cross}\le a,
\end{equation}
where $d_{\rm cross}$ is the distance from the cross–section center. Thus $p=0$ gives a uniform
``disk'' (no limb brightening) and $p>0$ yields an increasingly rim–dominated
``ring''. We adopt $p=1$ in the baseline and verify that the results are insensitive
within reasonable bounds.
In the azimuthally-averaged model limb-brightening enters
as a weight \(W_p(\xi)\), the LOS average of \((d_{\rm cross}/a)^p\)
along the same sightline (so \(W_p=1\) for \(p=0\) and $W_p$ increasing toward the rim
for \(p>0\)).

The final projected model is then
\begin{equation}
v_{\rm model}(x)
=\ -\,v_{{\rm LOS},\,{\rm wake}}
\ +\ v_\bullet \sin\theta\,\cos\theta\,\cos i\;
\big\langle \cos\phi \big\rangle(x)\; W_p(x),
\label{eq:final_model}
\end{equation}
for \(|x|<a\); for \(x\le -a\) (downstream plateau) we set
\(v_{\rm model}=-\,v_{{\rm LOS},\,{\rm wake}}\), and we do not compute
the model at \(x\ge a\) (i.e., upstream from the projected cross section).

\subsection{Fitting and Uncertainties}
\label{fit_kin.sec}

The fit parameters of the model are 
\[
\{\,v_\bullet,\ i,\ a,\ r_0\,\},
\]
the black hole velocity, the inclination, the projected half-width of the cross section,
and the projected position of the cross section along the wake. The radius of the
cross section is calculated from $i$ and $a$,
\[
R_{\rm ring}= \frac{a}{\sin i}.
\]
The following parameters can be varied but are held fixed in the fit:
\[
\{\,\chi,\ \theta,\ p\,\},
\]
the compression factor, the obliquity of the parabola, and the disk/ring geometry
of the cross section.
The sensitivity of the predicted velocities to the model parameters is
discussed in Appendix \ref{model_sens.sec}, where we also provide a simple
relation between the directly-observed velocity amplitudes and the
black hole's velocity and inclination.

We fit Eq.\ \ref{eq:final_model} to the observed velocity profile, where we combine
the [O\,III] and H$\alpha$ measurements at $r>61$\,kpc and include the Keck/LRIS
data of Paper I at $50<r<61$\,kpc to anchor the downstream component. We use
$\chi=3$ and $p=1$. For the obliquity we use $\theta =60^{\circ}$; we
show in \S\,\ref{compare_phot.sec} that this value is broadly consistent with
the morphology of the gas near the apex.\footnote{As $\Delta v_{\rm LOS} \propto \sin \theta \cos \theta = \frac{1}{2} \sin 2\theta$, $\Delta v_{\rm LOS}$ is maximal
for $\theta=45^{\circ}$.}
We minimize weighted
residuals $(v_{\rm model}-v)/\sigma_v$ with nonlinear least squares. Inside the
canopy we use Eq.~\ref{eq:final_model}; on the plateau ($x\le -a$) we use
Eq.~\ref{eq:wake_los}; $x\ge a$ is masked. Bounds enforce $v_\bullet\!\ge\!0$,
$0^\circ\!<\!i\!<\!90^\circ$, and $a_{\rm proj}\!\ge\!0$. 

\begin{figure}[ht]
  \begin{center}
  \includegraphics[width=0.95\linewidth]{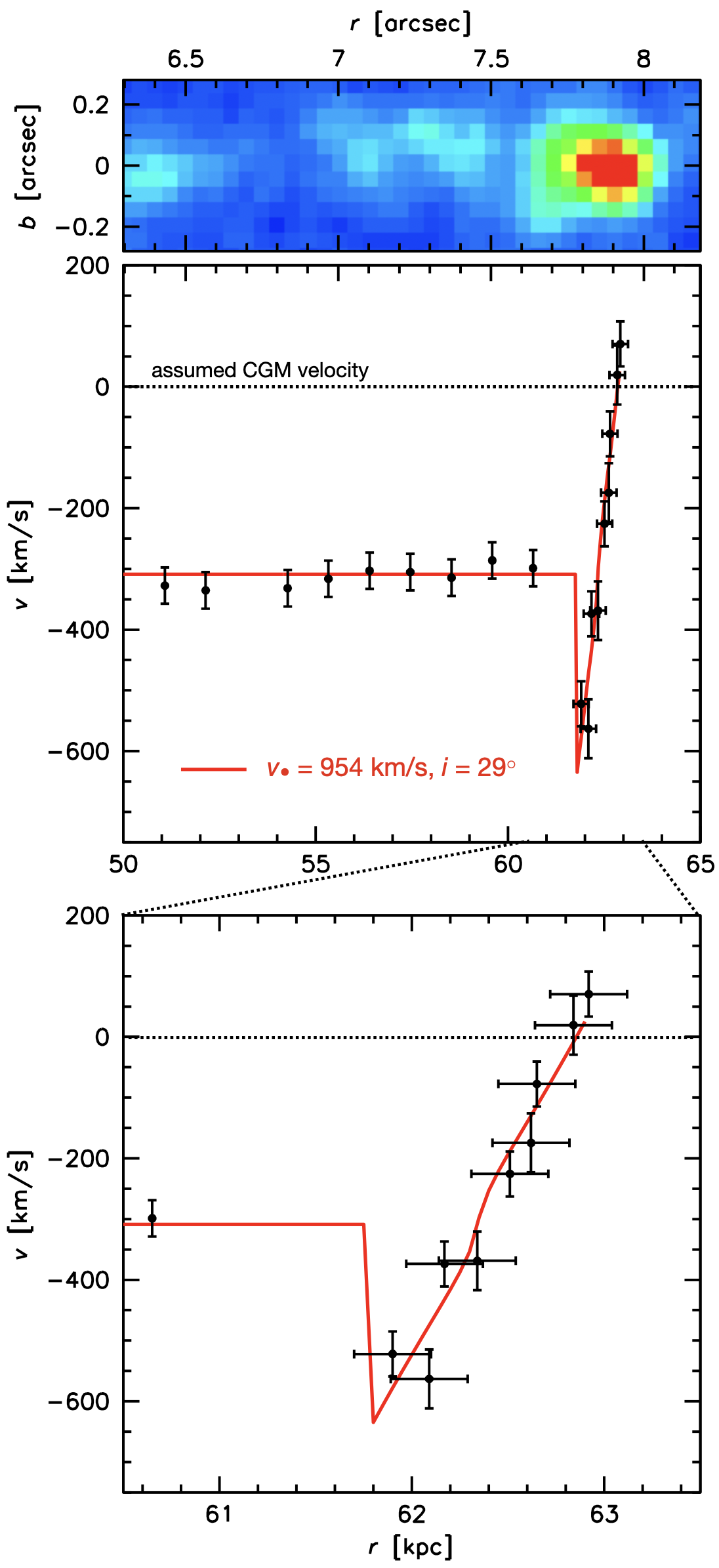}
  \end{center}
\vspace{-0.2cm}
    \caption{
Fit of our bow shock model to the observed kinematics near the apex (solid red
line).
The data are a combination of the [O\,III] and H$\alpha$ measurements of Fig.\
\ref{pv.fig} ($r>61$\,kpc)
and downstream measurements from the Keck/LRIS spectroscopy of Paper I ($r<61$\,kpc).
For reference the 
[O\,III] distribution is shown at the top.
The model provides an excellent fit to the data, with a reduced $\chi^2 \sim 1$,
as it
simultaneously explains the plateau velocity of $\approx -300$\,\kms\ and the
steep gradient at the tip. 
The assumed rest CGM velocity is $0$\,\kms.
}
\label{gradient_fit.fig}
\end{figure}

The best-fit profile is compared to the observations in Fig.\ \ref{gradient_fit.fig}.
The fit is good, with reduced $\chi^2 \sim 1$, and has
the following values for the key parameters:
\[
v_\bullet = 954^{+110}_{-126}\,{\rm km\,s}^{-1},\,
i=29^{+6}_{-3}\,{\rm deg},\
R_{\rm ring} = 1.5^{+0.7}_{-0.4}\,{\rm kpc}.
\]
The uncertainties are determined from simulations. 
The observations are perturbed by their uncertainties in both $r$ and $v$, and
the quoted confidence intervals are the 16$^{\rm th}$/84$^{\rm th}$
percentiles of the accepted samples. Note that the assumed
CGM rest velocity is $0$\,\kms, i.e., identical to the velocity of the
former host galaxy.

We conclude that the observed kinematics near the apex are consistent with a
bow shock produced by an object that is moving at $\sim 1000$\,\kms\
in a direction that is $\sim 30^{\circ}$ out of the plane
of the sky toward us.

\subsection{Comparison to the Photometric Analysis}
\label{compare_phot.sec}

Here we compare the kinematically-derived spatial extent of the bow shock to that
determined in \S\,\ref{morph.sec} from the parabola fit to the morphology
of the [O\,III]\,+\,H$\alpha$ emission.
The kinematics are weighted towards the location of the strongest line emission, approximately
2.0\,kpc downstream from the apex. Using the
best-fitting parabola of \S\,\ref{morph.sec} with $r_{\rm apex}\approx 64.6$\,kpc
and $R_{\rm c}\approx 1.8$\,kpc  we find a width of the shock
of $R_{\rm phot} \sim 2.7$\,kpc.\footnote{Note that projection
effects are minimal here, in contrast to the dynamical measurement. In the image plane there is foreshortening in the direction along the
wake $r$, but not -- to first order -- in the perpendicular direction $b$.}
This is somewhat larger than the cross-section radius that was determined
from our bow shock model fit to the kinematics, $R_{\rm kin} = R_{\rm ring}
= 1.5^{+0.7}_{-0.4}$\,kpc. The difference is likely caused by weighting: the emission
lines mostly arise in the dense compression zone interior to the shock surface.
The photometric
obliquity $\theta_{\rm phot}$ is related to $R_{\rm c}$ through $\theta_{\rm phot}
= \arctan\sqrt{2\Delta r/R_c}\approx 56^{\circ}$, similar to the assumed $\theta_{\rm kin}
= 60^{\circ}$.

We conclude that the spatial distribution of ionized gas at the tip of \wake\  and the
dynamical analysis are in reasonable agreement, particularly when the limitations of
our simplistic model are considered.

\begin{figure*}[ht]
  \begin{center}
  \includegraphics[width=0.92\linewidth]{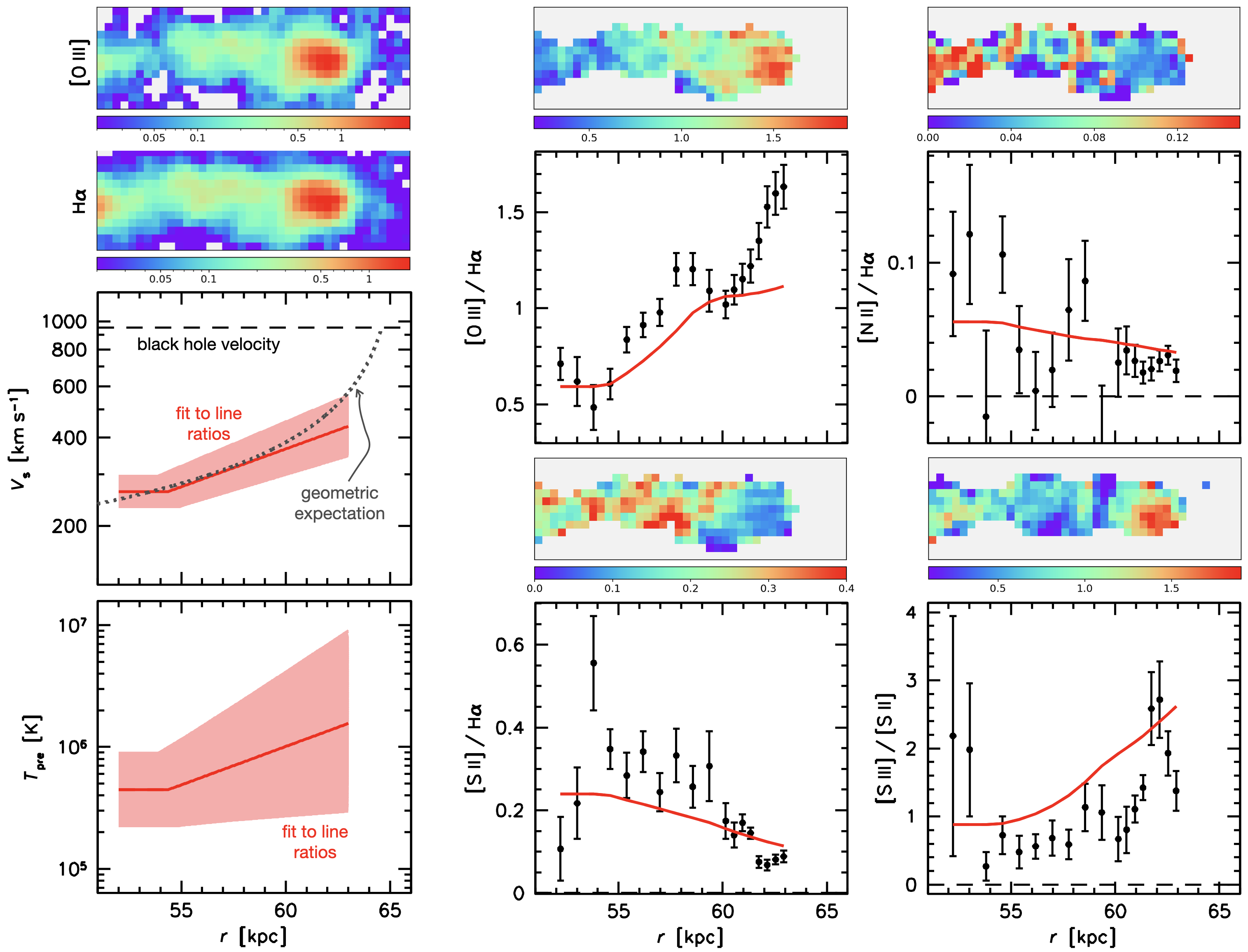}
  \end{center}
\vspace{-0.2cm}
    \caption{
{\em Left panels:} {\textsc Mappings} model variation in the shock velocity $V_{\rm s}$ and
preshock temperature $T_{\rm pre}$ that produces the fits shown at right.
The dotted grey curve shows the expected velocity variation based on the
kinematically-derived black hole velocity and the photometrically-derived
shape of the bow shock; it is in good agreement with the {\textsc Mappings} results.
{\em Middle and right panels:} Line ratio maps and line ratio
profiles along the wake. There is strong variation, with
[O\,III]/H$\alpha$ and [S\,III]/[S\,II]
increasing and [N\,II] and [S\,II]
decreasing toward the tip. Red lines show the best fit {\textsc Mappings} models.
They capture the trends, although there are significant local deviations.
}
\label{line_ratios.fig}
\end{figure*}

\section{Line Ratios}
\label{line_ratios.sec}

The JWST spectra cover several key diagnostic lines, ranging from [O\,III]\,$\lambda 5007$
to [S\,III]\,$\lambda 9532$. The strengths of these lines with respect
to H$\alpha$ provide independent constraints on the presence of a shock
and the velocity of the black hole. 

\subsection{Observed Line Ratios Near the Tip of the Wake}

Line fluxes are measured as a function of position along the wake ($r$),
after summing the flux in the perpendicular direction ($b$).
At $r>60$\,kpc, where the emission lines are relatively bright, the
wake is sampled
in individual $\Delta r = 0\farcs 05$ pixels. At $r<60$\,kpc we use
2-pixel ($\Delta r = 0\farcs 1$) bins to increase the S/N ratio.
Line fluxes are measured by summing the flux in wavelength bins
centered on the redshifted line, after subtracting the continuum. The continuum
is measured in an adjacent wavelength region; it is generally faint compared to
the line flux.

Line ratios are determined by dividing the calibrated line fluxes.
We consider [O\,III]/H$\alpha$, [N\,II]/H$\alpha$, [S\,II]/H$\alpha$, and
[S\,III]/[S\,II]. Here [O\,III] is [O\,III]\,$\lambda 5007$,
[N\,II] is [N\,II]\,$\lambda 6584$, [S\,II] is the sum of
the [S\,II]\,$\lambda\lambda 6716,6731$ doublet, and [S\,III] is the
sum of the [S\,III]\,$\lambda\lambda 9069,9532$ doublet.

The observed 1D line ratio profiles at $r>51$\,kpc are shown in Fig.\ \ref{line_ratios.fig},
along with corresponding 2D maps.
There are clear trends, with [O\,III]/H$\alpha$ rising from $\approx 0.6$ at $\approx 53$\,kpc
to $\approx 1.5$
at the apex; [N\,II]/H$\alpha$ low and decreasing slightly, reaching values of $\approx 0.02$ near
the apex; and [S\,II] and [S\,III] moving in opposite directions,
with [S\,II]/H$\alpha$ sharply decreasing and [S\,III]/[S\,II]
increasing with $r$.

\subsection{Line Ratio Modeling and Fitting}
\label{mappings.sec}

To interpret the observed emission line ratios along the wake we compare the measurements to a grid of fast radiative shock models generated with the \textsc{Mappings~V} code (Sutherland \& Dopita 2017, 2018).  We use the ``shock\,+\,precursor'' models, allowing the
relative weight of the
photoionizing precursor to be set with the parameter $f_{\rm p}$ (where $f_{\rm p}=1$ defaults to
weighting by the H$\alpha$ luminosity). In our main analysis we use $f_{\rm p}=1$.

Several other parameters are also held fixed in the fits, although we explore the
sensitivity to them.  
The preshock density is set to $n_{\rm pre}=1$, as
the relevant lines are formed when the density reaches $n\gg 1$. As expected, we find
that lowering the preshock density to more realistic CGM values ($n=0.01-0.1$) does not
change the results but leads to much longer computation times. 
For the metallicity we adopt SMC-like abundances ($Z \approx 0.2\,Z_\odot$), broadly consistent with results
from COS-Halos \citep[$Z_{\rm CGM} \sim 0.1-0.5\,Z_\odot$;][]{Tumlinson2017,Prochaska2017}. 
Increasing the metallicity leads to worse fits, particularly for [N\,II]/H$\alpha$.
 In the main analysis
we use $f_{\rm p}=1$, that is, the relative weight of the components is set by their
contribution to the $H\alpha$ luminosity.

We hold the magnetic parameter $B$ fixed to $B=1$ for our main analysis,\footnote{In \textsc{Mappings}, the magnetic parameter is \(B/\sqrt{n_{\rm pre}}\) with \(B\) in \(\mu\mathrm{G}\) and \(n_{\rm pre}\) in \(\mathrm{cm^{-3}}\); with our choice \(n_{\rm pre} = 1~\mathrm{cm^{-3}}\), \(B=1\) corresponds to a preshock field \(B_0 = 1~\mu\mathrm{G}\).}
but ran full grids of temperature and velocity with $B=0.3$ and $B=10$. In our fits
the magnetic parameter is degenerate with the preshock temperature (such that
lower $B$ leads to solutions with higher $T_{\rm pre}$) and the precursor weight $f_{\rm p}$.
The parameters $B$ and $f_{\rm p}$ 
are knobs that control the preshock ionization; changing them from their defaults
produces nearly identical fits but with a different form of $T_{\rm pre}(r)$.

The two parameters that we {\em do} allow to vary are the shock velocity $v_{\rm s}$ and the preshock
temperature (with the caveat that $T_{\rm pre}$ is sensitive to the choices for $B$ and $f$).
The model grid spans preshock temperatures
$T_{\mathrm{pre}} = 1\times 10^{2}$\,--\,$3\times10^{6}$~K and shock velocities
$v_{\mathrm{s}} = 50$\,--\,$2000~\mathrm{km~s^{-1}}$.  
Predicted fluxes for all relevant lines ([O\,\textsc{iii}]~$\lambda5007$,
[N\,\textsc{ii}]~$\lambda6584$, [S\,\textsc{ii}]~$\lambda\lambda6716,6731$, and
[S\,\textsc{iii}]~$\lambda\lambda9069,9532$) are normalized to H$\alpha$.

We construct two-dimensional interpolation surfaces 
$f_{i}(\log T_{\mathrm{pre}},\log v_{\mathrm{s}})$
for each diagnostic ratio $i$ using bicubic interpolation across the model grid.
For a given position along the wake ($r$), we parameterize the preshock temperature and shock velocity as
\begin{eqnarray}
\log T_{\mathrm{pre}}(r) & = &
\begin{cases}
\log T_{0}, & r < r_{\mathrm{t}},\\[4pt]
\log T_{0} + \alpha\,(r - r_{\mathrm{t}}), & r \ge r_{\mathrm{t}},
\end{cases}
\\
\log v_{\mathrm{s}}(r) & = &
\begin{cases}
\log v_{0}, & r < r_{\mathrm{t}},\\[4pt]
\log v_{0} + \beta\,(r - r_{\mathrm{t}}), & r \ge r_{\mathrm{t}},
\end{cases}
\end{eqnarray}
where $r_{\mathrm{t}}$ is a transition radius and $\{\,T_{0},v_{0},\alpha,\beta\,\}$
are free parameters.  The best fit is determined by minimizing
\begin{equation}
\chi^{2} = \sum_{j}\sum_{i}
\left[\frac{R_{i,j}^{\mathrm{obs}} - f_{i}(\log T_{\mathrm{pre}}(r_{j}),
\log v_{\mathrm{s}}(r_{j}))}{\sigma_{i,j}}\right]^{2},
\end{equation}
where $R_{i,j}^{\mathrm{obs}}$ and $\sigma_{i,j}$ are the observed ratios and their 1$\sigma$
uncertainties at position $r_{j}$.

We find $\log T_0 = 5.6 \pm 0.3$, $\log v_0 = 2.42 \pm 0.05$,
$r_t = 54 \pm 2$\,kpc, $\alpha = 0.06 \pm 0.05$, and $\beta = 0.026 \pm 0.005$.
The best fitting profiles for $T_{\mathrm{pre}}(r)$ and $v_{\mathrm{s}}(r)$,
as well as the accompanying spatial variations of the four fitted diagnostics,
are shown by the red lines in Fig.\ \ref{line_ratios.fig}.
In Appendix \ref{2dmaps.sec} the 1D profiles are converted to 2D maps, for
a direct comparison to the observed 2D line maps.
The fits reproduce the trends seen in the data reasonably well, although there are significant
deviations. These most likely stem from complexity in the morphology and ionization
structure of the wake, as is also evident from the spatial variations in the 2D
maps. In particular, near the tip of the wake there is a localized ``hot spot'' where both
[O\,III]/H$\alpha$ and [S\,III]/[S\,II] are elevated (see \S\,\ref{blackhole.sec} for further
discussion of this region). 

The preshock
temperature $T_{\rm pre}\sim 10^6$\,K, with considerable
uncertainty. This temperature is somewhat higher than expected
as optical line formation begins to be efficient
at $T \lesssim 10^5$\,K. As discussed above we find that increasing $f_{\rm p}$ or
$B$ brings $T_{\rm pre}$ down to this regime; rather than providing meaningful
specific values of $T_{\rm pre}$, the models indicate that strong
preshock ionization is required, and this can be achieved in several ways.

\subsection{Interpretation of the Shock Velocities}

The shock velocities are in the $v_{\rm s} = 250-450$\,\kms\ range, superficially
in  conflict with the kinematically-derived black hole velocity of $v_\bullet \sim 950$\,\kms.
However, they are in fact quite consistent.
For a bow shock, the jump conditions depend on the normal component of the flow, 
\begin{equation}
v_{\rm s}(\theta) = v_\bullet \cos \theta,
\end{equation}
with $\theta$ the angle between the symmetry axis and the local surface normal. This is
the same angle as the obliquity of the parabola discussed in \S\,\ref{morph.sec} and
\S\,\ref{vkin.sec}, and we can therefore
relate the expected shock velocity to the morphology
of the bow:
\begin{equation}
\label{vpred.eq}
v_{\rm s}(r) = v_\bullet \left( 1+\frac{2(r_{\rm apex} - r)}{R_c}\right)^{-0.5}.
\end{equation}
The curve for $v_\bullet = 954$\,\kms,
$r_{\rm apex}=64.6$\,kpc, and $R_c = 1.8$\,kpc is shown by the dotted line in
Fig.\ \ref{line_ratios.fig}.

The geometric expectation is in excellent agreement with the shock velocities
derived from the line ratios, particularly when considering that
the two measurements are completely independent:
the red curve 
is determined from emission line ratios, and the grey dotted curve reflects a combination
of a fit to the kinematics (providing $v_\bullet$) and a fit to the emission line
map of the shock (providing $r_{\rm apex}$ and $R_c$).

\section{Discussion}
\subsection{Origin of the Kiloparsec Scale Bow Shock}
\label{origin.sec}

The evidence for a supersonic bow shock at the head of \wake\ is very strong, bordering on
overwhelming. The spatial
distribution of the line emission, the velocity gradient, and the
\textsc{Mappings} analysis of the line ratios are all consistent with a bow shock associated with a compact object moving at
$\sim 1000$\,\kms\ through the CGM. The most direct evidence is the kinematic signature:
there is no other plausible explanation for a $\sim 600$\,\kms\ gradient over $0\farcs 1$.
Furthermore, the fact that the wake behind the shock points
toward the heart of a galaxy, combined with the 1000\,\kms\ velocity of the shock,
leave little doubt that the compact object in question is a runaway SMBH.

Here we discuss how the bow shock is produced. The most straightforward
explanation is the
interaction between gas that is gravitationally bound to the black hole and the
ambient CGM. However, the observed size of the bow far exceeds this
interaction scale.
The relevant scale for direct SMBH\,--\,gas interaction is the accretion radius,
\begin{equation}
  R_{\rm acc} \;=\; \frac{2GM_\bullet}{v_{\rm BH}^2 + c_s^2}
  \label{eq:Racc}
\end{equation}
\citep{BondiHoyle1944,Edgar2004}.  For \(M_\bullet \sim 2 \times 10^7\,M_\odot\),
\(v_\bullet\!\sim\!10^3\) km s\(^{-1}\), and hot CGM with $c_s \sim 200$\,\kms, Eq.~\ref{eq:Racc} gives
$R_{\rm acc} \sim 0.2$\,pc -- four orders of magnitude smaller than the observed kpc-scale
bow. The black hole's gravity is therefore not setting
the size or luminosity of the bow shock.

Instead, the shock is likely sustained by momentum balance. In
momentum-supported bows the stand-off distance is the location where the outward momentum flux in the gas around the body balances the incoming ram pressure:
\begin{equation}
  R_0^2 \;=\; \frac{\dot M_{\rm out} v_{\rm out}}{4\pi\,\rho_{\rm ext}\,v_\bullet^2}
  \label{eq:R0}
\end{equation}
(\citealt{Wilkin1996}; see also \citealt{vanBurenMcCray1988,CantoRagaWilkin1996}).
The spectacular bow shock of Mira \citep{Martin2007}
is a textbook case: a modest AGB wind speed of 5--10\,\kms\ but substantial
mass loss makes the product
$\dot M_{\rm out} v_{\rm out}$ large, pushing the bow
far from the star.\footnote{The
ratio of Mira's stand-off distance ($\approx 0.1$\,pc) to its
radius ($\approx 300$\,R$_{\odot}$) is a factor of $\sim 10^4$.}
Simulations have successfully
reproduced both the observed stand-off scale and the wake morphology of Mira
in this framework \citep{Martin2007,Wareing2007,Esquivel2010}.

In our case, the outward pressure is probably not supplied
by a true wind coming from the SMBH, at least not consistently.
Inverting Eq.~\ref{eq:R0} and setting $\rho_{\rm ext} = \mu m_p n_{\rm ext}$, the 
outward mass flow required for a kpc-scale stand-off radius is 
\begin{equation}
\label{eq:mout}
\dot M_{\rm out} \approx 
\frac{v_\bullet^2}{v_{\rm out}} 4 \pi R_0^2\,\mu m_p n_{\rm ext} \approx 0.26\,M_{\odot}\,{\rm yr}^{-1}\,\left(\frac{10^3\,{\rm km}\,{\rm s}^{-1}}{v_{\rm out}}\right)
\end{equation}
for $\mu=0.62$, $n_{\rm ext}=10^{-3}$\,cm$^{-3}$, $v_\bullet = 954$\,\kms, and
$R_0=1.2$\,kpc
(see \S\,\ref{morph.sec}). 
Even a strong SMBH-driven wind with $v_{\rm out} \sim 3000$\,\kms\ requires a mass
outflow rate of
$\sim 0.1$\,\msun\,yr$^{-1}$, corresponding to $\sim 10^7$\,\msun\ since the
SMBH left its former host galaxy. Escaping black holes can carry
mini–disks with them, but their maximum mass is only a few
$\times10^{6}\,M_\odot$ for $v_\bullet \sim10^3$\,\kms\ and 
$M_\bullet\sim 2\times10^{7}\,M_\odot$
\citep{Loeb2007PRL}. Therefore, if the main source of outward pressure
were a SMBH-driven wind, it would have been exhausted long ago.

\begin{figure}[ht]
  \begin{center}
  \includegraphics[width=0.95\linewidth]{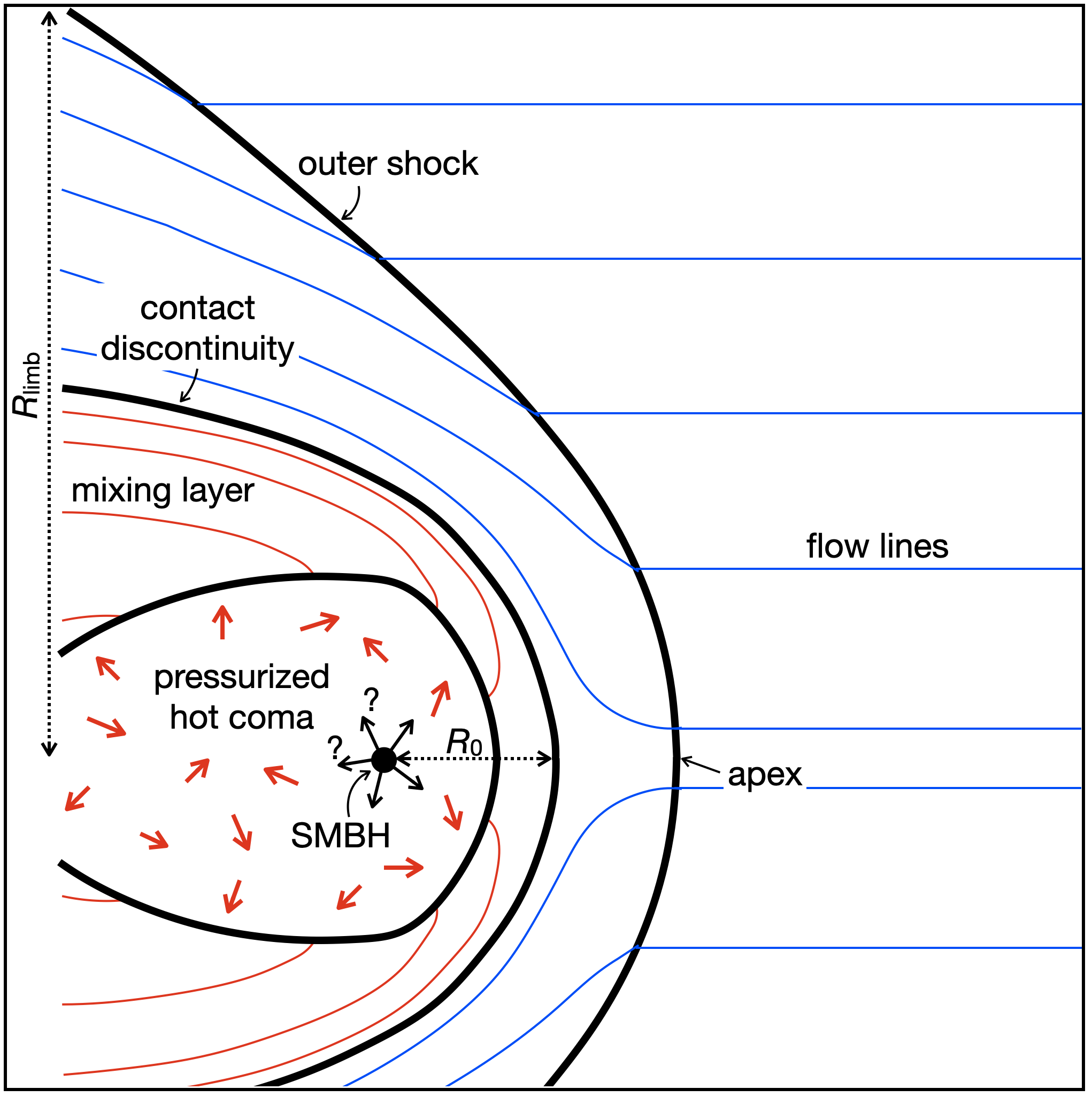}
  \end{center}
\vspace{-0.2cm}
    \caption{
Illustration of the
momentum-balanced bow shock that we propose for \wake. The figure is
adapted from \citet{tarango:18}, replacing their stellar wind by a pressure-supported
coma. The CGM produces an inward velocity that is captured over the limb
cross section $R_{\rm limb}$. This wind is balanced at the contact
discontinuity by pressure from a hot, mass-loaded coma
with radius $R_0$. Activity associated with the SMBH may provide additional
outward pressure.
}
\label{coma.fig}
\end{figure}

Momentum balance does not require a true wind, as it can also be achieved by
an effective wind generated by a pressurized, mass-loaded coma.
This ``windless''  geometry is illustrated
in Fig.\ \ref{coma.fig} and is analogous to cluster cold fronts with bow shocks 
such as Abell 3667 \citep{Vikhlinin2000A3667} and the bullet cluster \citep{Markevitch2002Bullet}. 
In this picture the relevant outward velocity
is the post-shock sound speed of the hot gas at the apex. For
a strong  adiabatic shock $c_s \approx 0.56 v_s$, with $v_s$ the
local shock velocity \citep{SutherlandDopita2017}.\footnote{This relation is not given explicitly in \citet{SutherlandDopita2017} but it follows from their
expression for the post-shock temperature, $T_{\rm ps} = \frac{3\,\mu m_p}{16\,k}\,v_s^2$,
combined with $c_s = \sqrt{\frac{\gamma kT}{\mu m_p}}$, with $\gamma = 5/3$ for
an ideal monatomic gas.}
The shock velocities near the apex are
$v_s = 500\text{--}700$\,\kms\ (see Eq.\ \ref{vpred.eq}), giving an effective
outflowing wind of $v_{\rm out} \approx c_s \approx 300\text{--}400$\,\kms\ and
(using Eq.\ \ref{eq:mout}) a mass flow of $\approx 0.7$\,\msun\,yr$^{-1}$.

The source for this mass is the intercepted
CGM, and we can quantify this as
\begin{equation}
  \dot M_{\rm in}\;\approx\;\rho_{\rm ext} v_\bullet \pi R_{\rm limb}^2,
  \label{eq:mdot_in}
\end{equation}
where $R_{\rm limb}$ is the effective cross section of the limbs.
As Eqs.~\ref{eq:R0} and \ref{eq:mdot_in} both scale with the density, the condition
$\dot M_{\rm in} \gtrsim \dot M_{\rm out}$ is satisfied if
\begin{equation}
\label{eq:balance}
\frac{R_{\rm limb}}{R_0} \gtrsim \sqrt{\frac{4v_\bullet}{v_{\rm out}}}.
\end{equation}
For our fiducial values of $v_\bullet=954$\,\kms, $v_{\rm out}=350$\,\kms,
and $R_0=1.2$\,kpc, this implies an effective
limb radius of $R_{\rm limb} \gtrsim 4$\,kpc, similar
to the cross section of the 
bow at the location of peak emission ($R_{\rm phot} \sim 2.7$\,kpc; see 
\S\,\ref{compare_phot.sec}).

In summary, 
the bow is maintained by the thermal pressure of the shocked and recycled
CGM at the apex, and likely persists in a quasi-steady state for
tens of Myr. Continuous inflow and shock heating of the external gas
replenishes the hot, mass-loaded coma, which provides the pressure needed to
balance the upstream ram pressure. 
The resulting kiloparsec scale of
the shock -- far exceeding the SMBH sphere of influence  -- is set by the momentum budget
\citep[see, e.g.,][]{Wilkin1996,MarkevitchVikhlinin2007}. No 
wind source beyond the CGM
is required to sustain the bow, although it is possible that
activity associated with the SMBH
contributes to $v_{\rm out}$.
The geometry of the shock surface is governed by Eq.\ \ref{eq:balance}; if, for example,
black hole activity temporarily boosts $v_{\rm out}$, the apex becomes overpressured
and the stand-off distance $R_0$ increases in response.

\subsection{Origin of the Wake}

The feature that initially drew our attention to \wake\ is not the bow shock but the spectacular wake behind it. Here we discuss the origin of the wake and
reinterpret the downstream [O\,III] velocity trend that we reported in Paper I.

\subsubsection{Entrainment and Mixing with the CGM}
\label{downstream.sec}

\begin{figure}[ht]
  \begin{center}
  \includegraphics[width=0.95\linewidth]{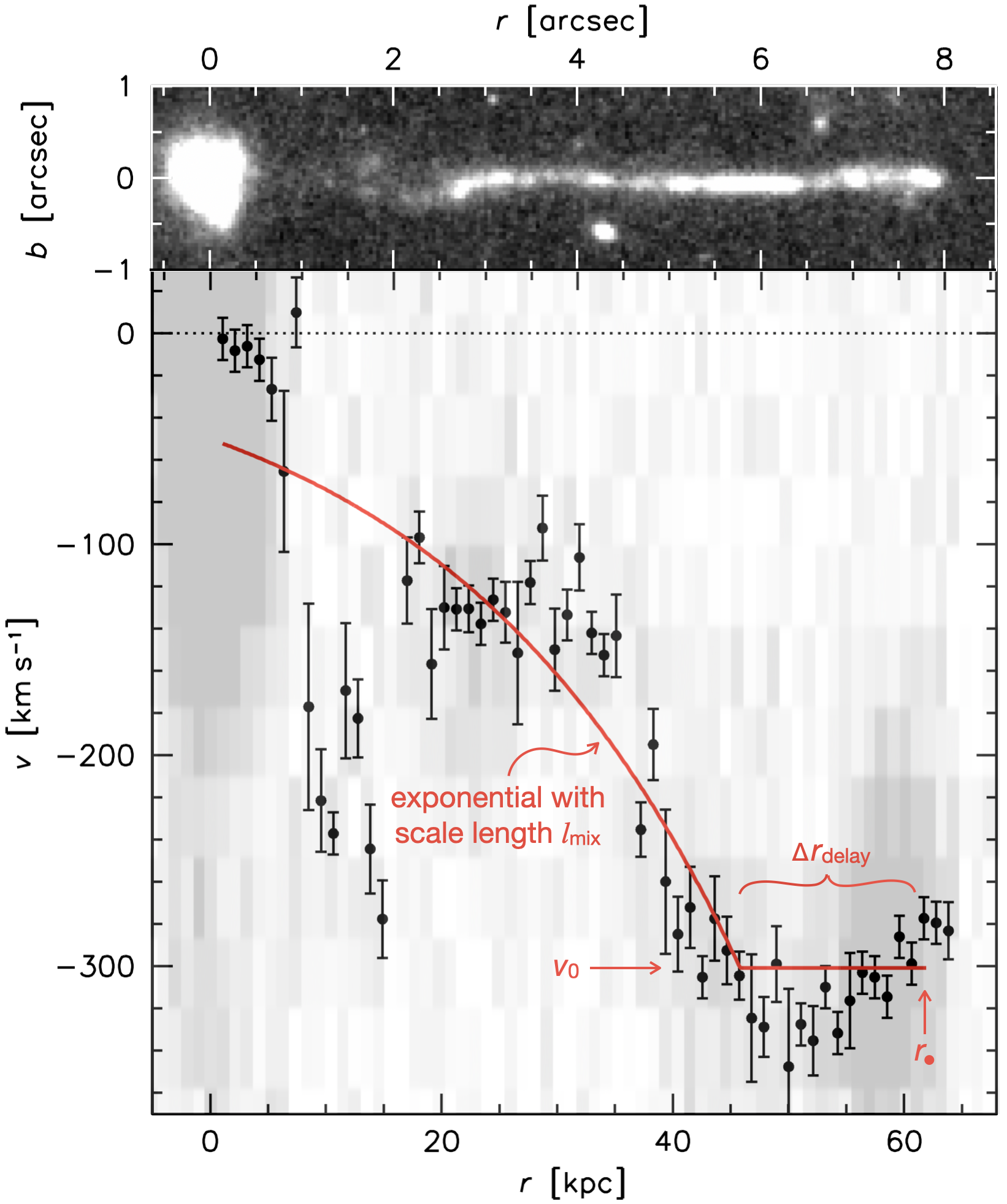}
  \end{center}
\vspace{-0.2cm}
    \caption{
Velocity measurements along the wake, from ground-based [O\,III] spectroscopy
described in Paper I. The 2D Keck/LRIS spectrum is shown in light grey.
The HST UVIS F200LP\,+\,F350LP white light image is shown at the top for reference.
The curve is a delayed-mixing model, with key parameters indicated.
The model has a
delay length of $\Delta r_{\rm delay} \sim 16$\,kpc and an exponential
mixing length of $l_{\rm mix} \sim 26$\,kpc. These
values are broadly consistent with mixing models of
cold gas flows in the CGM (see text).
}
\label{downstream.fig}
\end{figure}

The wake is highly complex. The emission line gas 
just behind the bow shock is fragmented into several distinct clumps
(see Fig.\ \ref{apex.fig}). Ground-based spectroscopy shows
that line emission
is present  along the entire extent of the wake, with some regions showing
clear evidence for ongoing star formation and others exhibiting [O\,III]/H$\beta$
ratios indicative of shocks (see Paper I). 
The radial velocity of the gas shows a strong gradient, going
from $\sim -300$\,\kms\ near the tip 
to $\sim -100$\,\kms\ at $r\sim 20$\,kpc, where $v=0$\,\kms\
is the radial velocity
of the galaxy (see Paper I and Fig.\ \ref{downstream.fig}).
Despite all these turbulent processes
the wake remains
strikingly narrow over most of its 62\,kpc extent, with a radius of $R_{\rm wake}
\approx 0.7$\,kpc.\footnote{We set the wake radius as $0.5\times$ its FWHM in the HST/UVIS image.} It is also dynamically cold, at least near
the tip where we can measure it, with
a velocity dispersion of $31 \pm 4$\,\kms\ (see Appendix \ref{dispersion.sec}).

All these features -- the coherence of the wake over many kpc,
the presence of cold gas far downstream, the evidence for shocks
and star formation, and the velocity gradient -- are hallmarks of 
radiative mixing with the CGM.
When a thin cold gas stream flows in a hot medium
a turbulent mixing layer is established between the cold and the hot
gas, with the ambient gas becoming entrained and condensing out of the
mixing layer. This allows for a long-lived, $T\sim10^4$\,K phase to persist along the wake,
with the cold gas mass growing with time \citep[see][]{gronke:2018,ji:natast2019,
gronke:2022}. As the newly added gas has $\langle v \rangle \approx 0$\,\kms,
the mean absolute velocity of the gas in the wake decreases with time.
Where the condensed gas becomes sufficiently dense it can form stars
in situ, as has been seen in simulations of stripped
tails \citep{tonnesen:2012}. These processes are
not smooth: in the entrainment framework the
turbulence in the mixing layer is often supersonic with respect to the
$10^4$\,K phase, explaining the localized internal shocks and
high-excitation patches that are seen within the  wake
\citep{scannapieco:2015,mandelker:2016}.

The observed velocity gradient can be compared to predictions
for turbulent radiative mixing layers \citep[see, in particular,][for
the analogous case of infalling cold gas onto galaxies]{tan:23}.
Efficient mixing is expected to start after a delay
$\Delta t_{\rm delay}$, as it takes time for the Kelvin-Helmholtz instabilities
to grow and cooling of the mixing layer to become efficient.
After that, the cold gas mass is expected to grow
exponentially, as the added mass depends
on the amount of mass that is already present. Momentum conservation implies that
$\dot{v}/v = -\dot{m}/m$, and therefore we have $v(t) = v_0 \exp(-t/t_{\rm grow})$ with
$v_0$ the initial velocity and $t_{\rm grow}$ defined as $t_{\rm grow} \equiv m/\dot{m}$.

Motivated by these considerations, and making use of the fact that time
corresponds to position along the wake,
we fit the following empirical model to the observed velocity gradient:
\begin{equation}
v(r) = v_0 
\exp \left(-\frac{\max\{0,\,(r_\bullet-r)-\Delta r_{\rm delay}\}}{l_{\rm mix}}\right),
\label{eq:delayed}
\end{equation}
with $r$ the distance from the galaxy along the wake,
$r_\bullet \approx 62$\,kpc the current  position of the black hole,
$v_0$ the plateau velocity immediately behind the black hole,
$\Delta r_{\rm delay}$ the
distance behind the black hole where mixing becomes important, and
$l_{\rm mix}$ the mixing length. All length scales are in the observed
frame, and hence projected.
The best fit is shown in Fig.\ \ref{downstream.fig}. The model
provides a reasonable description of the data, although there
are significant deviations (particularly in the region $r=10-20$\,kpc,
where the wake also loses morphological coherence).
We find  $v_{0} = -301\pm 6$\,\kms,
$\Delta r_{\rm delay} = 16 \pm 2$\,kpc, and
$l_{\rm mix} = 26\pm 4$\,kpc.

Interpreting these values in terms of time rather than projected distance,
we have $\Delta t_{\rm delay} =
\Delta r_{\rm delay} / (v_\bullet\,\cos\,i) \sim 19$\,Myr, and 
$t_{\rm grow} = l_{\rm mix} / (v_\bullet\,\cos\,i) \sim 31$\,Myr.
These timescales are consistent with the results of \citet{tan:23} for
radiative mixing of cold gas clouds in a hot ambient medium. Specifically,
they find $t_{\rm grow} \sim 35$\,Myr for typical CGM conditions (their
Eq.\ 26). The age of the wake is $t_{\rm wake} = r_\bullet / (v_\bullet\,\cos\,i) \sim 73$\,Myr,
and we conclude that, for the
oldest parts of the wake closest to the galaxy, entrainment
has increased the cold gas mass by a factor of
\begin{equation}
\frac{M}{M_0} \sim \exp \left(\frac{t_{\rm wake} - \Delta t_{\rm delay}}{t_{\rm grow}}\right) \sim 6.
\end{equation}

\subsubsection{Mass Budget: Can Entrainment Explain the Stellar Mass of the Wake?}

We now ask whether entrainment provides sufficient cold gas to explain the observed
luminosity of the wake.
The wake has an F814W AB magnitude of $\approx 22.9$ (Paper I). 
Transforming this into a rest-frame $B$ band luminosity gives $L_B \approx 2 \times
10^{10}\,L_{\odot,B}$, and assuming that most of the light comes from  stars
with ages $1-20$\,Myr and $(M/L)_B \sim 0.01$, we find a total present-day
stellar mass of
$M_* \sim 2 \times 10^8$\,\msun. Taking stellar mass loss
into account, the total formed mass was $M_{*,0} \approx M_*/0.7 \sim 3
\times 10^8$\,\msun.

The CGM mass that is intercepted by the bow is $\dot{M}_{\rm in}
\approx 0.7$\,\msun\,yr$^{-1}$ for $R_{\rm limb} \approx 4$\,kpc
(see \S\,\ref{origin.sec}). Since the SMBH left the galaxy the
total intercepted mass is $M_{\rm bow} \approx t_{\rm wake} \dot{M}_{\rm in}
\approx 5\times 10^7$\,\msun. Taking $M_{\rm bow}$ as the seed for entrainment
$M_0$, we find a total entrained mass of order $M_{\rm ent} \sim 6\,M_{\rm bow}
\sim 3\times 10^8$\,\msun.

While the total cold gas mass and the stellar mass appear to be in good agreement,
this is only the case for an unreasonably efficient 
gas\,$\rightarrow$\,stars conversion. The highest conversion rates occur in starburst
galaxies, which have gas depletion times of $t_{\rm dep} = 0.2\pm 0.1$\,Gyr
\citep{genzel:10}. Even with this depletion time, the maximum 
star formation efficiency over the $\approx 70$\,Myr lifetime
of the wake would be $\epsilon_{\rm max} \approx 1 - \exp(-70/200) \approx 0.3$,
with $\int_t M_* = \epsilon \int_t M_{\rm gas}$.
The stellar mass that can form out of the entrained gas is
therefore $M_* \lesssim 0.3 M_{\rm ent} \lesssim 1\times 10^8$\,\msun.
We conclude that the
stellar mass that can  be produced by the entrained gas is at least
a factor of several lower than the observed stellar mass, for our
default assumptions.

There are several ways to resolve this mild discrepancy,
such as a higher CGM density (as $\dot{M}_{\rm in} \propto \rho_{\rm ext}$)
or a larger cross section of the bow.
An interesting possibility is that the stellar initial mass function
(IMF) is modestly top-heavy, thus lowering the $M/L$ ratio of the wake.
A relatively high background temperature may raise the turnover
mass of the IMF due to the $T^{3/2}$ dependence of the Jeans mass
\citep{larson:98,krumholz:06,bate:09,bate:23}. Top-heavy
IMFs have been observed in some other extreme environments, such
as the 30 Doradus cluster \citep{schneider:18} and the central
300\,pc of the Milky Way \citep{lu:13,hosek:19}.

\subsection{Black Hole Mass Estimate from Energy Conservation}

It is perhaps surprising that the black hole mass does not enter the analysis
of \S\,\ref{origin.sec}: the expression for the geometry of the bow shock
depends on the velocity of the SMBH and the effective counteracting
wind only. The SMBH
plays an important role behind the scenes: its gravity keeps a few-pc
core bound and centered, preventing the stagnation point from wandering and
thereby stabilizing the coma and bow-shock geometry.
The mass likely also mattered during the initial growth phase, as gas that escaped
with the black hole may have seeded the bow. Ablation
and mixing of this gas with the CGM pressurized the coma
until momentum balance with the ram pressure
was reached. Once that happened, the bow became self-sustaining. 

Although we cannot directly measure the mass of the SMBH from the
observed shock, 
the fact that the shock has persisted for $t_{\rm wake}\sim 70$\,Myr provides an interesting
constraint.  
The bow continuously thermalizes the kinetic energy of the
CGM at a rate similar to the ram–pressure power \citep[e.g.,][]{Vikhlinin2001,
MarkevitchVikhlinin2007}:
\begin{equation}
  \dot E_{\rm heat} \;\sim\; \rho_{\rm ext}\,v_\bullet^{3}\,\pi R_{0}^{2}.
  \label{eq:ram_power}
\end{equation}
Over the wake lifetime $t_{\rm wake}$, the total processed kinetic energy
is therefore 
\begin{equation}
  E_{\rm heat} \;\sim\; \varepsilon\,\rho_{\rm ext}\,v_\bullet^{3}\,
  \pi R_{0}^{2}\,t_{\rm wake},
\end{equation}
where $\varepsilon\!\sim\!1$ is the fraction of ram power deposited into
the hot phase.  The ultimate source of this heating is
the kinetic energy of the SMBH, and we have
$E_\bullet = \tfrac12 M_\bullet v_\bullet^{2} \gtrsim E_{\rm heat}$. Rewriting
this in terms of a mass limit, we obtain
\begin{equation}
  M_\bullet \;\gtrsim\;
  2\,\varepsilon\,
  \rho_{\rm ext}\,v_\bullet\,\pi R_{0}^{2}\,t_{\rm wake}.
  \label{eq:MBH_lower_limit}
\end{equation}
For our fiducial values
$n_{\rm ext}= 10^{-3}\,{\rm cm^{-3}}$, $v_\bullet=954~{\rm km\,s^{-1}}$,
$R_{0}=1.2~{\rm kpc}$, and $t_{\rm wake}=73~{\rm Myr}$, 
Eq.~\ref{eq:MBH_lower_limit} gives  
$M_\bullet \gtrsim {\rm a\,\,few}\times 10^{7}\,M_\odot$.

Interestingly, this mass is similar to the SMBH mass of $M_\bullet \sim 2 \times 10^7$\,\msun\
that is implied by the bulge mass of the former
host galaxy (see Paper I and \citealt{schutte:19}).
This similarity implies that
the SMBH may have slowed down somewhat since its escape. Furthermore, it suggests that
gravitational wave recoil is a more likely escape
mechanism than a three-body interaction. Recoil causes the merged remnant black hole
to escape, leaving a black hole-less galaxy behind.
In three-body interactions only one of the three SMBHs generally
escapes, and it is the lowest mass SMBH of the three that has the highest
escape probability
\citep[e.g.,][]{hoffman:07}. In this scenario it is unlikely
that the mass of the escaped SMBH is similar to the
{\em total} SMBH mass that is implied by the
bulge mass\,--\,$M_\bullet$ scaling relation.

\subsection{Can We See the Black Hole?}
\label{blackhole.sec}

All observational signatures that we have discussed in this paper so far are caused
by the bow shock and not by the SMBH itself.\footnote{The black hole is the Wizard
behind the curtain: it is the unseen mover setting the boundary conditions, but
what we actually observe is the shocked CGM.} 
We already knew that the black hole
is not very active: in Paper I we examined publicly available Chandra and VLA data,
finding no significant signal near the tip of the wake. 
When we wrote the JWST proposal we expressed hope that
we might detect the SMBH even if it is not active,
through gas that is bound to it; in that case
we would see a point source with relatively
broad (${\rm FWHM} \gtrsim 500$\,\kms) emission lines. A careful search
of the data cube near the tip does not show
such an object. Moreover, it may not be possible
to detect bound gas even if it exists:
in the geometry of Fig.\ \ref{coma.fig} the SMBH is inside a hot
($T \gtrsim 10^6$\,K) coma that suppresses optical line formation.

We do find tentative evidence for emission originating from the SMBH, not in
the JWST spectra but in the HST/UVIS LP$_{\rm diff}$ UV image. As explained
in \S\,\ref{uvis.sec} the LP$_{\rm diff}$ band is obtained by subtracting
two long pass filters, ${\rm LP}_{\rm diff} = {\rm F200LP} - 1.22\times
{\rm F350LP}$, creating a broad UV filter with central rest-frame wavelength
$\lambda_{\rm rest} \sim 1400$\,\AA. It is the most sensitive UV filter
on HST but it is not very ``clean'',
as the F200LP and F350LP bands are not identical beyond 3500\,\AA. as a result,
LP$_{\rm diff}$ contains varying amounts of light from long wavelengths
depending on the color of the object, and is almost impossible to calibrate.

\begin{figure}[ht]
  \begin{center}
  \includegraphics[width=1.0\linewidth]{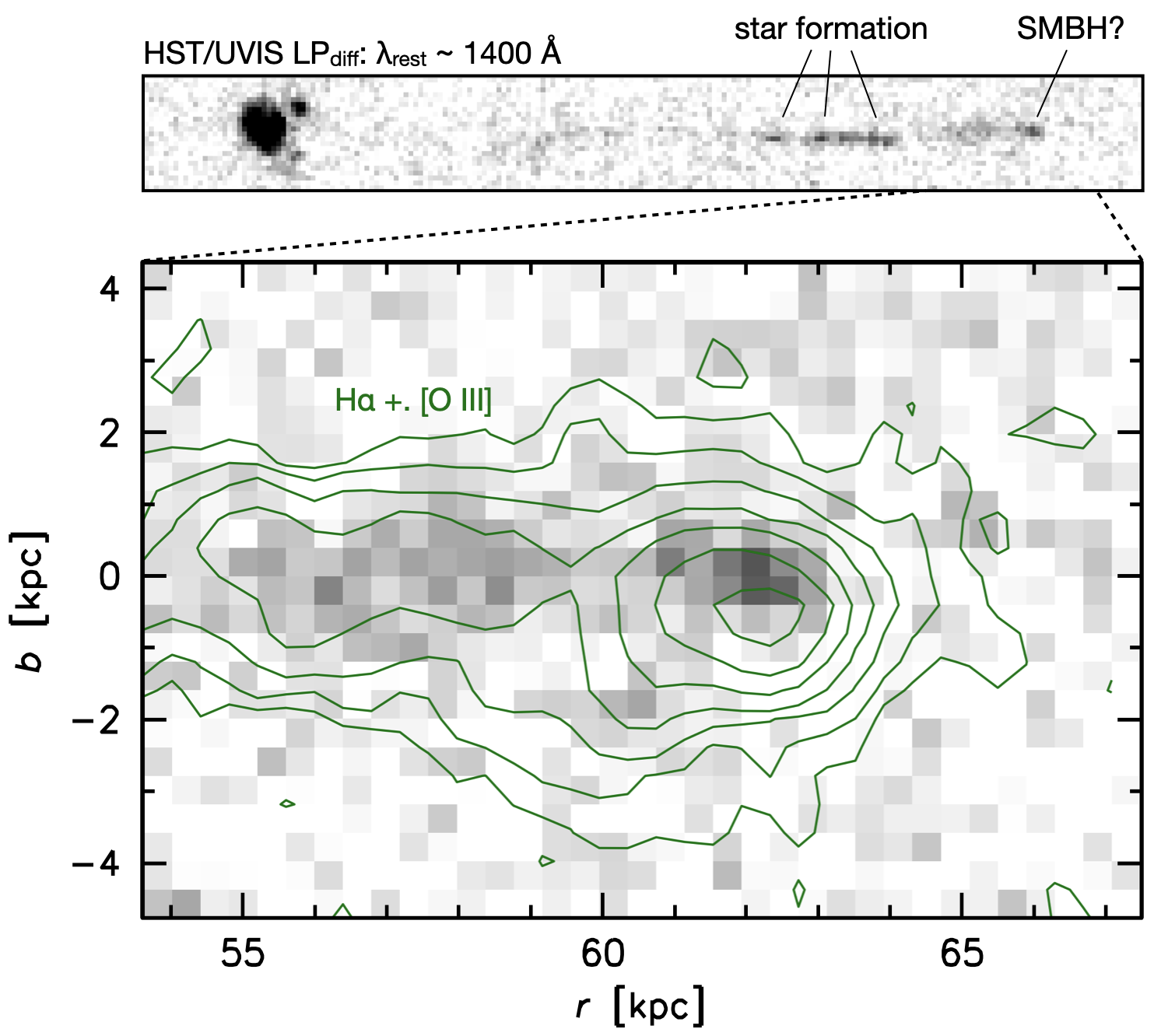}
  \end{center}
\vspace{-0.2cm}
    \caption{
{\em Top panel:} HST/UVIS far-UV image of \wake, in the broad LP$_{\rm diff}$
passband. The wake shows several downstream knots that are associated
with star forming regions (see Paper I). There is a compact UV source near
the tip of the wake, which may be emission from the vicinity of the SMBH.
{\em Main panel:} Zoom-in on the tip region, with contours
showing the [O\,III]\,+\,H$\alpha$ line emission. The UV knot is in
the expected location for the SMBH, $1-2$\,kpc behind the apex.
It is slightly offset
from the peak of the line emission.
}
\label{uv.fig}
\end{figure}

With these caveats in mind, we show the LP$_{\rm diff}$ image in the
top panel of Fig.\ \ref{uv.fig}. The wake shows regions of diffuse emission,
particularly just behind the tip, and several distinct knots. The 
interpretation of the downstream
knots is straightforward: they coincide with star forming regions that we identified
in Paper I based on their (ground-based) emission line ratios.

There is also a knot at the tip of the wake. This may be a star forming region
as well, but the line ratios in that vicinity are inconsistent with that
interpretation (see \S\,\ref{line_ratios.sec}). A zoomed-in view
is shown
in the main panel
of Fig.\ \ref{uv.fig}.
Intriguingly, while the UV knot is offset from the peak
of the line emission, it is in the expected location 
for the SMBH: on the symmetry axis, at a distance $R_0 = 1-2$\,kpc behind the apex of
the shock surface (see Fig.\ \ref{coma.fig}).
We tentatively identify the UV source as emission
from the vicinity of the SMBH itself, although this has to be confirmed
with follow-up UV imaging or spectroscopy.

We note that the line ratio analysis provides 
circumstantial support for this interpretation.
As discussed in \S\ref{line_ratios.sec}, there is a small region near the tip
where the $[$O\,III$]/\mathrm{H}\alpha$ and $[$S\,III$]/[$S\,II$]$ ratios are elevated. In that patch the observed $[$O\,III$]/\mathrm{H}\alpha$ exceeds the values from shock $+$ precursor models 
\citep{SutherlandDopita2017}, indicating an additional source of hard ionizing photons.
This patch is $0\farcs 1$ from the UV source and it could be that photons
from the SMBH illuminate the inside of the mixing layer at that location.
Further circumstantial support comes from the fact
that even weakly-active SMBHs can launch large-scale winds \citep[see, e.g.,][]{cheung:16,
cazzoli:22},
thus making it easier to balance the external ram pressure
(see \S\,\ref{origin.sec}).

\subsection{Revisiting Paper I}

With the additional information from the HST/UVIS imaging
and the JWST/NIRSpec spectroscopy
we briefly discuss what aspects of the analysis in Paper I are confirmed
or refuted by the new data.

The central proposal of Paper I was that the linear feature is the wake behind
a runaway supermassive black hole, and this is strongly supported by our analysis. We also confirm the
presence of a spatially-resolved bow shock at the head of the wake, something that
we predicted based on shock models
and the luminosity of the [O\,III] knot in the Keck/LRIS data.

What we did {\em not} get right is the interpretation of the [O\,III] velocity
curve. We assumed that the CGM remains unperturbed by the wake except for small
perpendicular gravitational accelerations, as in the \citet{delafuente:08} and
\citet{ogiya:24} models, and it did not occur to us to interpret the ionized
gas as a post-shock flow with velocity $\approx (1-1/r)v_\bullet$.
Instead, we suggested that ionizing
radiation from the shock and the wake illuminates the CGM and that the data probe
the local CGM kinematics. We did realize this was problematic,
as the velocity trend exceeds the estimated virial velocity
of the halo ($V_{\rm vir}\approx 130$\,\kms).\footnote{We also realized
that the CGM had to be dynamically cold in our explanation; as shown in
Appendix \ref{dispersion.sec} the velocity dispersion of the
gas is only $\approx 30$\,\kms. We therefore suggested that the wake
was illuminating a cold filament in the CGM.}
Our interpretation of ``wiggles'' in the morphology
as evidence of buffeting by the CGM is also wrong (\S\,7.2 in Paper I). There
is no need for a complex explanation; in the
radiative cooling\,/\,entrainment picture local clumping and displacements
are expected, due to the turbulent nature of the process.

Importantly, we find no evidence for a proposed ``counter'' wake, on the other side
of the galaxy. We tentatively detected this feature in [O\,III]
and a ground-based CFHT $u$ band image.  If the CFHT detection
were real, the feature should have been quite bright in our deep
HST/UVIS F200LP and F350LP data, but it is undetected. This was also reported
by \citet{montes:24}, who performed an archival study of the UVIS data.
The proposed counter wake is also undetected in the deep VLT $B$ band imaging
discussed in \citet{dokkum:23rnaas}. The lack of a counter feature
resolves an awkward coincidence that we highlighted in Paper I:
dual escapes of SMBHs are possible but rare
compared to single escapes, occuring in only $\sim 1$\,\% of three-body
interactions \citep{hoffman:07}.

Finally, the stellar population
analysis of \S\,7.1 may or may not be correct. We interpreted
the strong color variation along the wake as alternating dominance of blue
and red supergiants, with stellar population ages that
increase monotonically with distance from the black hole. That interpretation
likely holds qualitatively, but the distance\,$\rightarrow$\,age
model that we fitted is
probably too simplistic. In the entrainment picture, star formation 
not only occurs just behind the tip but downstream as well, as new
gas is continuously added to the wake. We see direct evidence for this
in the UV image. A broad age gradient is
still expected, as the black hole passage sets a maximum age of 
$(r_\bullet-r)/(v_\bullet\,\cos i)$ at any position along the wake.
A re-analysis of the continuum emission, making use of the new HST
imaging and JWST spectra, is possible
but beyond the scope of the present paper.

\section{Conclusions}

Using newly obtained HST/UVIS and JWST/NIRSpec data we confirm that the
remarkable linear feature reported in Paper I is the wake behind a runaway
SMBH. The single strongest piece of evidence is a $\sim 600$\,\kms\
velocity gradient over $0\farcs 1$ at the tip of the feature. We dub the
object \wake, recognizing that it is the first confirmed runaway SMBH.
\wake\ is an emperical validation of
the 50-year old prediction that SMBHs can escape from
their host galaxies, through 
gravitational wave recoil \citep{bekenstein:73}
or a three-body interaction \citep{saslaw:74}.

Wakes may be a generic feature of runaway black holes, as all that
may be needed is
a small initial gas reservoir to seed the coma and bow. If so,
their demographics can directly constrain the frequency and velocity
distribution of SMBH ejections.
\wake\ is difficult to identify in ground-based imaging
data as its extreme axis ratio of $\approx 50$ is reduced by an order
of magnitude in typical seeing conditions. The obvious data sets to look for
these features in a systematic way are wide-field surveys with Euclid and Roman.

\begin{acknowledgements}
We thank Daisuke Nagai and Go Ogiya for useful discussions in the early stages
of this project. The comments from the anonymous referee improved the clarity and
presentation of the results.
Support from grants HST-GO-17301 and JWST-GO-3149 is gratefully acknowledged.
The data from these programs  can be retrieved using DOI
\dataset[10.17909/8d8q-r813]{https://doi.org/10.17909/8d8q-r813}.
\end{acknowledgements}

\bibliography{master_0925}{}
\bibliographystyle{aasjournal}

\begin{appendix}

\section{The Velocity Gradient Interpreted as the Gravitational Influence of the Black Hole}
\label{grav.sec}

In the main text we interpret the steep velocity gradient near the apex in the context
of a bow shock model. Here we ask whether the gradient could be caused by the
gravitational influence of the SMBH --  either from gas that is bound to the SMBH and escaped
with it \citep[see][]{boylankolchin:04,merritt:09}, or from kicks imparted on
the gas as the black hole passes \citep[see][]{ogiya:24}.

We find that the spatial extent of the gradient is at least
an order of magnitude too large. We express this in terms of the
derived black hole mass, which can be compared to the $M_\bullet \sim 2\times 10^7$\,\msun\
mass that was derived from the $M_\bullet$\,--\,$M_{\rm stars}$ relation of
\citet{schutte:19} in Paper I.
Turning first to the case of
gravitationally-bound gas, the observed $\sim 600$\,\kms\ velocity
gradient implies an orbital speed of $v\approx 300$\,\kms\
at $r\approx 1$\,kpc. The implied mass is
\begin{equation}
M_\bullet \approx \frac{v^2 r}{G}
\approx 2\times 10^{10}\ M_\odot,
\end{equation}
higher by three orders of magnitude than expected from the stellar mass of the galaxy.

One could also imagine that the gradient arises from hyperbolic encounters of the SMBH with ambient gas, as explored by \citet{ogiya:24}.
In the Ogiya \& Nagai impulsive limit, the parallel kick at impact parameter $b$ is
\begin{equation}
\Delta v_{\parallel} \approx \frac{2\,G^{2}M_\bullet^{2}}{V^{3}\,b^{2}}.
\end{equation}
Conservatively
interpreting the observed $\sim 600\ \mathrm{km\,s^{-1}}$ gradient as
a symmetric two–sided shear gives $M_\bullet \approx 1\times10^{11}\ M_\odot$
for $\Delta v_{\parallel}\approx300\ \mathrm{km\,s^{-1}}$, $b\approx 1$\,kpc,
and $V\approx 1000$\,\kms.

We infer that in either interpretation an implausibly massive SMBH is required,
and therefore reject a purely gravitational origin
(bound or impulsive) for the gradient.

\section{Sensitivity of the Kinematics Fit to Model Parameters}
\label{model_sens.sec}

\subsection{Effect of Parameter Variations}

\begin{figure*}[ht]
  \begin{center}
  \includegraphics[width=1.0\linewidth]{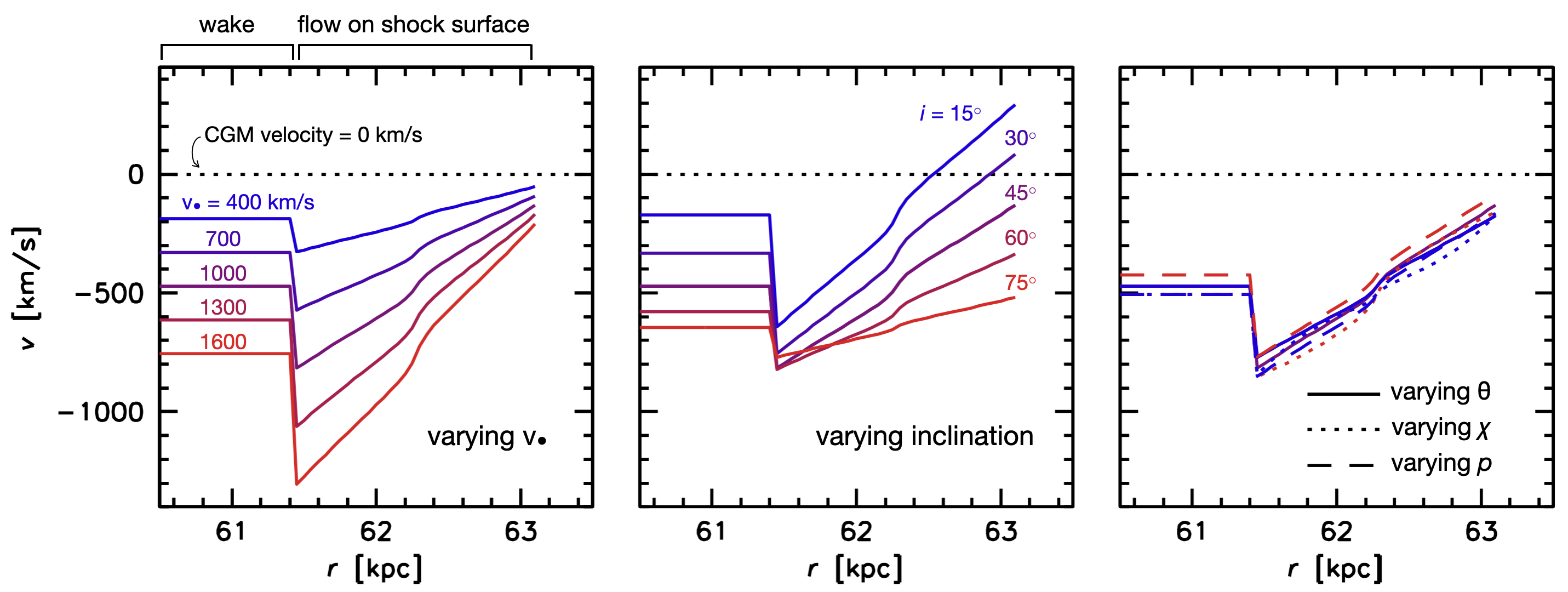}
  \end{center}
\vspace{-0.2cm}
    \caption{
Dependence of the velocity profile on the parameters of the model. The default
model here has black hole velocity
$v_\bullet = 1000$\,\kms, inclination $i=45^{\circ}$, obliquity $\theta=60^{\circ}$,
compression factor $\chi=3$, and disk/ring parameter $p=1$. 
{\em Left:} Effect of varying $v_\bullet$. {\em Middle:} Effect of varying the
inclination. {\em Right:} Effect of varying $45^{\circ}\leq \theta \leq 70^{\circ}$,
$2.5 \leq \chi \leq 3.5$, and $0.5 \leq p \leq 2$. The black hole mass and inclination
are the dominant parameters, and they can be determined independently from the combination
of the plateau velocity and the amplitude of the gradient.
}
\label{model_variations.fig}
\end{figure*}

In \S\,\ref{fit_kin.sec} 
we modeled the velocity flows across the bow shock to constrain
the black hole velocity $v_\bullet$, the inclination $i$,
and the radius of the cross section $R_{\rm ring}$.
In Fig.\ \ref{model_variations.fig} we show the effects of varying the
fit parameters. The black hole velocity,
shown at left, is effectively a scaling of the entire curve. 
The inclination has a qualitatively different effect, as it impacts
the ratio between the amplitude of the gradient ($\propto \cos i$)
and the plateau velocity ($\propto \sin i$). The black hole velocity and
the inclination can therefore be inferred from the observed
velocity curve (see below).

The right panel of Fig.\ \ref{model_variations.fig} shows the effect of changing
the ``fixed'' parameters $\chi$ ($2.5 \leq \chi \leq 3.5$),
$\theta$ ($30^{\circ} \leq \theta \leq 60^{\circ}$, and $p$ ($0.5\leq p \leq 2$).
The impact on the velocity profiles is small compared to that of the black hole
velocity and the inclination.

\subsection{A Simple Method to Estimate the Black Hole Velocity and Inclination}
\label{simple.sec}

Here we point out that $v_\bullet$ and $i$
can be estimated directly from the observed position\,--\,velocity diagram.
Assuming that the amplitude of the
gradient $v_{\rm grad}
= (v_{\rm max}-v_{\rm min})/2$ and the absolute plateau velocity $|v_{\rm plat}|$ have
been measured, and ignoring the azimuthal and
disk/ring factors in Eq.\ \ref{eq:final_model}, the inclination can be inferred 
directly from $|v_{\rm plat}|/v_{\rm grad}$:
\begin{equation}
i \approx \arctan\left[ \frac{\sin 2\theta}{2(1-1/r)} \frac{|v_{\rm plat}|}
{v_{\rm grad}}\right].
\end{equation}
For  $\theta \approx 60^{\circ}$ and $\chi \approx 3$ this reduces to
\begin{equation}
i \approx \arctan \left( 0.65\frac{|v_{\rm plat}|}{v_{\rm grad}}\right).
\end{equation}
The black hole velocity follows from the inclination and the plateau velocity:
\begin{equation}
v_\bullet = \frac{\chi}{\chi-1} \frac{|v_{\rm plat}|}{\sin i}
\approx \frac{3}{2} \frac{|v_{\rm plat}|}{\sin i}
\end{equation}
for $\chi\approx 3$. 
Applied to our data, we have $v_{\rm grad} \approx 317$\,\kms,
$|v_{\rm plat}|\approx313$\,\kms, and therefore
$i \approx 33^{\circ}$ and $v_\bullet \approx 860$\,\kms.
These values are only $\sim 10$\,\% removed from the results
of the full model fit. While a full fit is obviously more accurate, the
approximation given here may be useful for the interpretation of future
kinematic data on other supersonic shocks.

\section{2D Maps from 1D Fits to Line Ratios}
\label{2dmaps.sec}

\begin{figure}[ht]
  \begin{center}
  \includegraphics[width=0.65\linewidth]{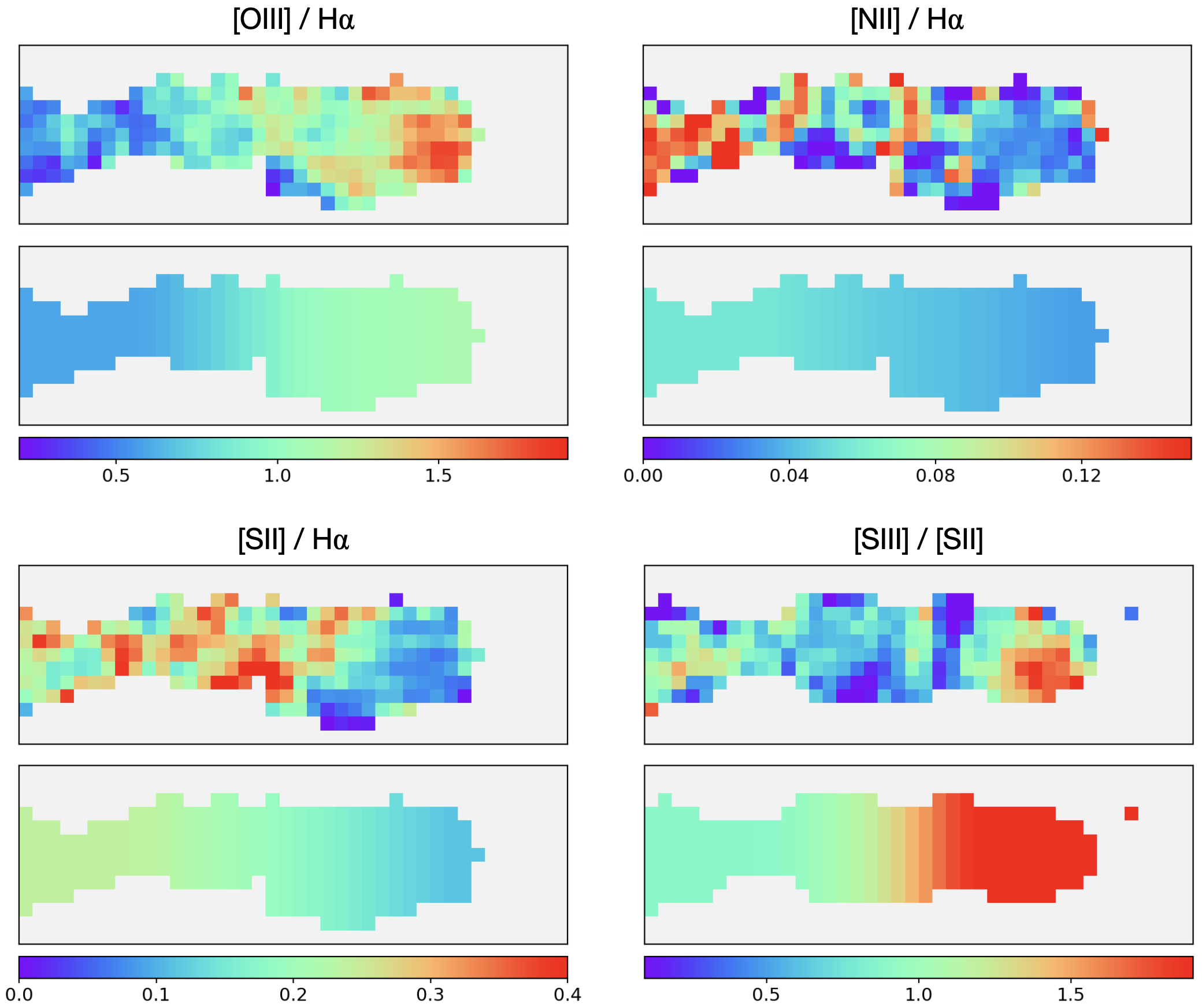}
  \end{center}
\vspace{-0.2cm}
    \caption{
Comparison of observed 2D line ratio maps to 2D models, created from the
1D model fits of \S\,\ref{line_ratios.sec}. The trends are reasonably well
reproduced but there are local deviations, particularly in a knot
near the tip. The overall normalization
of [S\,III]/[S\,II] is too high in the model.
}
\label{model_2d.fig}
\end{figure}

In the analysis of \S\,\ref{line_ratios.sec} the observed 2D line ratio maps are
collapsed in the $b$ direction, perpendicular to the wake. These 1D observed profiles
are fit with shock models, producing the profiles shown by
the red lines in Fig.\ \ref{line_ratios.fig}. Here we replicate the best-fitting
1D profiles in the $b$ direction so they can be visually compared to the observed
2D maps.  The observed and model maps are shown in Fig.\ \ref{model_2d.fig},
using identical color scaling and masking for each of the line ratios.

As noted in \S\,\ref{line_ratios.sec} and \S\,\ref{blackhole.sec}
the main deviations for [O\,III]/H$\alpha$ are in a compact
knot near the very tip, possibly
due to additional ionization from the black hole itself. The 2D profiles
of [N\,II/H$\alpha$] and [S\,II/H$\alpha$] are reasonably well-reproduced by
the 1D model. In [S\,III]/[S\,II] the same knot is seen as in [O\,III/H$\alpha$],
but here the overall normalization of the model is higher than the observed
ratios.

\section{Velocity Dispersion Measurement from LRIS Data}
\label{dispersion.sec}

\begin{figure}[ht]
  \begin{center}
  \includegraphics[width=0.45\linewidth]{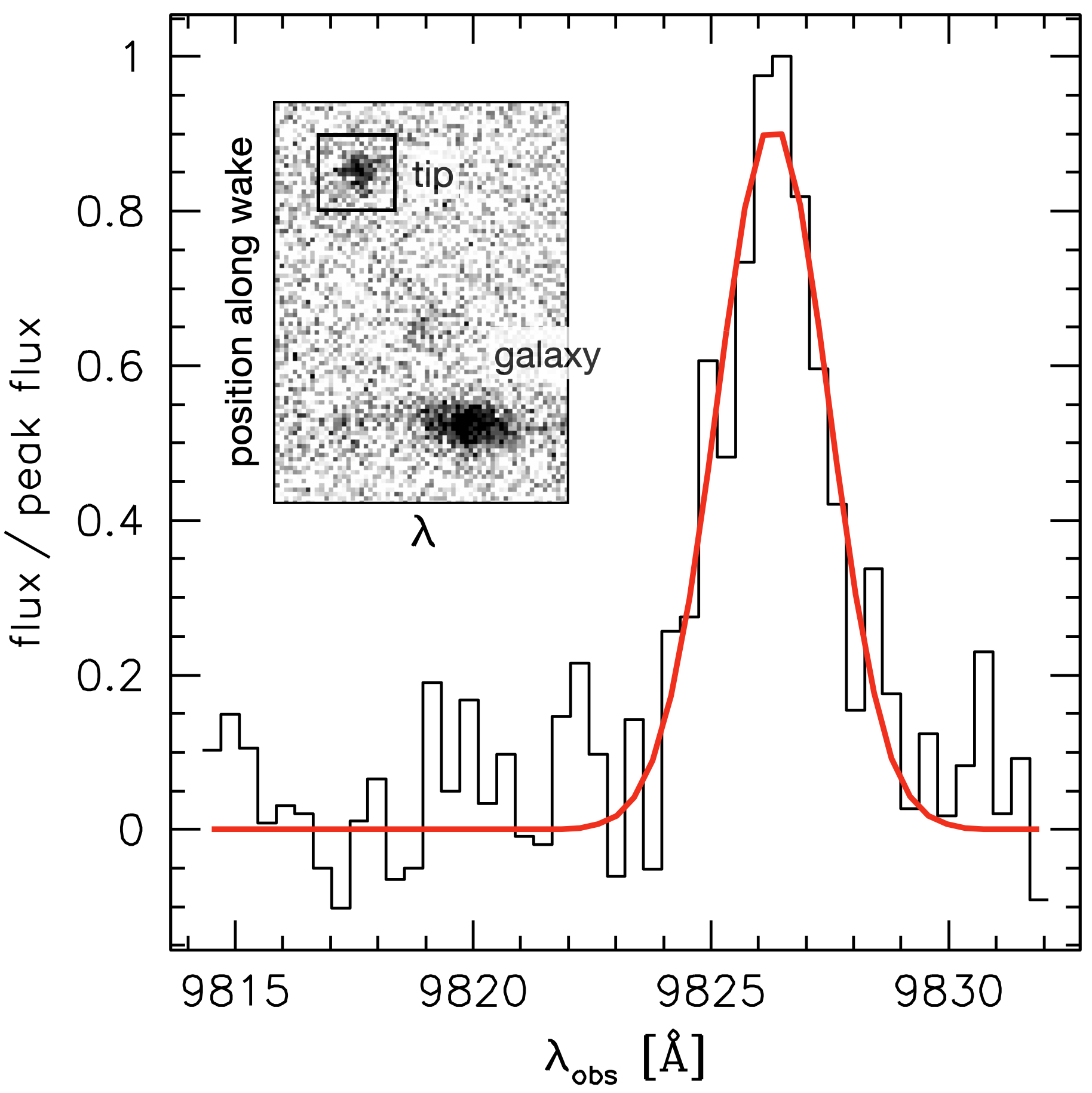}
  \end{center}
\vspace{-0.2cm}
    \caption{
{\em Inset:} High resolution 2D Keck/LRIS spectrum of \wake\
in the [O\,III]\,$\lambda 5007$ region, with
$\sigma_{\rm instr}=18$\,\kms. These data were previously shown in Fig.\ 1 of Paper I.
{\em Main panel:} Velocity profile of the knot at the tip of the wake,
with the best-fitting Gaussian in red. The width of the line is $\sigma=30\pm 4$\,\kms\
after correcting for instrumental resolution.
}
\label{dispersion.fig}
\end{figure}

The analysis in Paper I
was largely based on low resolution spectra obtained with the 400\,lines\,mm$^{-1}$
grating, as these have the highest S/N ratio. However,
we also obtained higher resolution observations with the 1200\,lines\,mm$^{-1}$ grating.
The spectral resolution (as measured from sky emission lines) is $\sigma_{\rm instr}=18$\,\kms.
The 2D spectrum near the [O\,III]\,$\lambda 5007$ line is shown in
Fig.\ 1 of paper I, and repeated in the inset of
Fig.\ \ref{dispersion.fig}.
The main panel of Fig.\ \ref{dispersion.fig} shows the averaged spectrum
of the [O\,III] knot at the tip. A Gaussian fit gives $\sigma_{\rm obs}=
1.17 \pm 0.14$\,\AA, or $36\pm 4$\,\kms. Correcting for the instrumental resolution gives
$\sigma =  31 \pm 4$\,\kms\ for the dispersion of the gas.

The dispersion is significantly higher than the thermal broadening: at $T \sim 10^4$\,K the thermal dispersion is $\sigma_{\rm th}\approx 2{-}3~\mathrm{km\,s^{-1}}$ for oxygen ions. 
The two components that are likely contributing to the dispersion are the $\sim 600$\,\kms\
velocity gradient induced by the bow shock limbs and turbulence.
Disentangling these components requires data of higher S/N ratio and spatial resolution.

\end{appendix}

\end{document}